\documentclass[%
 reprint,
 superscriptaddress,
 amsmath,amssymb,
 aps,
]{revtex4-2}
\usepackage{amssymb}
\usepackage{graphicx}
\usepackage{dcolumn}
\usepackage{booktabs}
\usepackage{textcomp}
\usepackage{bm}
\usepackage{xcolor}
\usepackage{hyperref}

\usepackage{color}
\usepackage{cleveref}

\begin{document}

\title{Scalable Generative Sampling and Multilevel Estimation for Lattice Field Theories Near Criticality}

\author{Ankur Singha}
\email{a.singha@tu-berlin.de}
\affiliation{Berlin Institute for the Foundations of Learning and Data (BIFOLD), Germany}
\affiliation{Technische Universit\"{a}t Berlin, Germany}
\author{Jacob Kauffmann}%
\affiliation{Berlin Institute for the Foundations of Learning and Data (BIFOLD), Germany}
\affiliation{Technische Universit\"{a}t Berlin, Germany}

\author{Elia Cellini}%
\affiliation{Higgs Centre for Theoretical Physics, School of Physics and Astronomy, The University of Edinburgh, Edinburgh EH9 3FD, United Kingdom}

\author{Karl Jansen}
\affiliation{Deutsches Elektronen-Synchrotron (DESY), Platanenallee 6, 15738 Zeuthen, Germany}
\affiliation{Computation-Based Science and Technology Research Center, The Cyprus Institute, Nicosia, Cyprus}


\author{Shinichi Nakajima}
\email{nakajima@tu-berlin.de}
\affiliation{Berlin Institute for the Foundations of Learning and Data (BIFOLD), Germany}
\affiliation{Technische Universit\"{a}t Berlin, Germany}
\affiliation{RIKEN Center for Advanced Intelligence Project (AIP), Japan}
\date{\today}

\begin{abstract}
   Sampling lattice field theories near criticality is severely hindered by critical slowing down, which makes standard Markov chain methods increasingly inefficient at large lattice volumes. We introduce a multiscale generative sampler, inspired by renormalization-group ideas, that models the Boltzmann distribution through a coarse-to-fine hierarchy across length scales. At each level, a conditional Gaussian mixture model captures the main local dependence of newly introduced variables on the already-sampled coarse field, while a masked continuous normalizing flow refines the remaining conditional structure. Coarse levels encode the dominant long-wavelength modes, and finer levels progressively add short-distance fluctuations. In addition, because the architecture preserves coarse fields exactly during refinement, it provides exact restriction maps at no additional computational cost and directly enables asymptotically unbiased Multilevel Monte Carlo (MLMC) variance reduction. For the two-dimensional scalar $\phi^4$ theory near criticality, the method achieves integrated autocorrelation times orders of magnitude smaller than Hybrid Monte Carlo (HMC) on large volumes, maintains high importance-sampling efficiency relative to other generative baselines, and reproduces asymptotically unbiased physical observables in statistical agreement with long HMC simulations.

\end{abstract}
\maketitle

\section{Introduction}
Markov Chain Monte Carlo (MCMC) simulations are the standard framework for sampling lattice field theories and computing observables from first principles~\cite{WilsonKogut1974,creutz1983quarks,CREUTZ1983201}. Their impact is particularly evident in lattice QCD, where large-scale simulations have enabled precision determinations of hadronic observables~\cite{FLAG:2024oxs}, as well as first-principles calculations of hadronic contributions to the muon anomalous magnetic moment~\cite{Aoyama:2020ynm,Aliberti:2025beg}. At the same time, the reliability and scope of these calculations depend directly on the availability of algorithms capable of generating representative field configurations efficiently across a wide range of couplings, lattice spacings, and volumes.

A major limitation of MCMC methods is the rapid increase of computational cost near criticality, and more generally as the continuum limit is approached. As the correlation length grows in lattice units, slow long-wavelength modes come to dominate the dynamics and conventional Markov-chain algorithms suffer from severe critical slowing down~\cite{wolff1990critical,wolff2004monte,Schaefer:2010hu}. Quantitatively, the integrated autocorrelation time typically scales as a power of the correlation length, $ \tau_{\mathrm{int}} \sim \xi^z $, so the cost of producing statistically independent configurations rises rapidly with increasing system size and decreasing lattice spacing. In some theories, this problem is further compounded by the slow evolution of global or topological modes. Overcoming these dynamical bottlenecks is therefore essential for extending precision lattice calculations to larger volumes and finer lattices.

Motivated by these limitations, and by the rapid progress of deep learning methods in high-dimensional generative modeling, alternative sampling strategies based on neural networks have attracted increasing attention in lattice field theory. The central idea is to learn a parametric approximation of the target Boltzmann measure, thereby enabling the generation of proposals with low autocorrelation. Among the corresponding deep generative models, Normalizing Flows (NFs)~\cite{Tabak:2010,tabak2013family,rezende2015variational,chen2018neural} have emerged as a particularly prominent framework. These algorithms have been explored in several forms, including coupling-layer flows~\cite{PhysRevD.100.034515,Kanwar:2020xzo,albergoflowbased,nicoli_pre,nicoli_prl,Singha:2023cql,Singha:2023xxq,Abbott:2024kfc,tuo2025scalable,Abbott:2026ylv,Faraz:2023xdi,Finkenrath:2024pR,Kreit:2026eng,Schuh:2026dvp,10.21468/SciPostPhys.16.5.132,faraz2023improvement}, continuous normalizing flows~\cite{Bacchio:2022vje,gerdes2023learning,caselle2024sampling,Gerdes:2024rjk}, and stochastic normalizing flows~\cite{Caselle:2022acb,Caselle:2024ent,Bulgarelli:2024yrz,Bulgarelli:2024brv,bonanno2026scalable}. Other generative approaches have also been studied, notably variational autoregressive networks~\cite{PhysRevLett.122.080602,Singha2025RiGCS,BIALAS2022108502,Bialas:2025hxu,du2026scalingautoregressivemodelslattice}, diffusion models~\cite{Wang:2024xyz,Zhu:2025pmw,aarts2026generalizableequivariantdiffusionmodels,Kanwar:2025wuc,vega2025group,Kanaujia:2025jru,Alharazin:2026lcb} and GANs~\cite{Pawlowski_2020,Singh_2021,10.21468/SciPostPhysCore.5.4.052}.

However, their practical usefulness for large-scale lattice simulations is ultimately limited by scalability~\cite{DelDebbio:2021qwf,Abbott:2022zsh}. This limitation is particularly severe near criticality for single-scale architectures, where the model must learn the joint distribution of the entire lattice at a single resolution. Capturing the long-range correlations near criticality typically requires large receptive fields or very deep networks, leading to a rapid increase in training cost, memory usage, and sampling inefficiency near criticality as the system size grows.

Several strategies have been proposed to mitigate these issues. Among them, stochastic normalizing flows can improve scalability by introducing auxiliary non-equilibrium MCMC updates and effectively constructing non-equilibrium transport paths between simple and target distributions~\cite{Bulgarelli:2024brv,bonanno2026scalable}. While this stochasticity can significantly enhance sampling efficiency at large volumes, it comes at the cost of requiring a predefined annealing schedule and a more involved generation procedure.

Normalizing flows can nevertheless be highly effective in specific applications. A notable example is the calculations of derivatives of physical observables, where flow-based changes of variables in the path integral can be optimized to reduce the variance of the corresponding estimators~\cite{Abbott:2024kfc,Catumba:2025ljd,Abbott:2026ylv}. However, this advantage is tied to the optimization of specific observables and does not directly translate into scalable, fully generative samplers of the full lattice distribution. The tension between expressivity, scalability, and generation cost therefore remains a central challenge for their application to large-scale lattice field theory.

A natural route towards improving scalability is to exploit the multiscale structure of critical lattice systems. Under coarse graining, the slow infrared modes of the fine lattice are transferred to coarser levels, where they appear as shorter-wavelength modes in lattice units and can therefore be treated more efficiently. This idea underlies a long history of multiscale methods in lattice field theory and Monte Carlo simulations, including early multigrid Monte Carlo approaches~\cite{Goodman:1986pv,Goodman:1989jw,Janke:1993ut} and simulation techniques inspired by renormalization-groups~\cite{Schmidt:1983rng,FAAS1986571,Endres:2015yca,PhysRevD.102.114512}. In multigrid Monte Carlo, long-distance modes of the fine lattice are represented and updated on coarser grids, after which the resulting coarse-grid corrections are interpolated back to the fine lattice as collective updates. This strategy can significantly reduce critical slowing down in controlled settings, such as free or weakly interacting theories. However, it remains an acceleration method for fine-lattice MCMC rather than a direct coarse-to-fine generative scheme, and its performance can deteriorate when efficient coarse-grid updates become difficult to construct for the underlying theory~\cite{PhysRevD.47.R3103} and also in the presence of a nontrivial global or topological structure~\cite{Goodman:1989jw}.

Renormalization-group-inspired approaches~\cite{Schmidt:1983rng,FAAS1986571,Endres:2015yca,PhysRevD.102.114512}, by contrast, proceed more directly in a coarse-to-fine direction, aiming to reconstruct fine configurations by sequentially introducing short-distance fluctuations from coarse variables. However, this requires renormalized effective densities at each scale, which are generally intractable beyond simple models. Deep generative models offer a way around this limitation by learning such effective distributions implicitly, without requiring their explicit construction. This idea has been successfully leveraged in Ref.~\cite{Singha2025RiGCS}, where variational autoregressive networks were used to model the coarse scale of the two-dimensional Ising model, leading to an efficient deep generative sampler with reduced training and sampling costs relative to single-scale models. Similar machine learning approaches have also been investigated in Refs.~\cite{BIALAS2022108502,Bialas:2022bdl,Abbott:2024knk,bauer2025super,ihssen2025generativesamplingphysicsinformedkernels,du2026scalingautoregressivemodelslattice}.

In this work, we generalize the multilevel generative strategy of Ref.~\cite{Singha2025RiGCS}, originally developed for discrete systems, to continuous lattice field theories. We introduce a multiscale generative sampling framework in which the Boltzmann measure is represented through a hierarchy of coarse-to-fine conditional densities such that long-range correlations are captured at coarse levels and short-distance fluctuations are introduced progressively during refinement. At each refinement step, a conditional Gaussian mixture model provides a tractable local approximation to the conditional distribution of the newly introduced fine degrees of freedom given the coarse field, and a continuous normalizing flow then refines this approximation toward the target conditional density. The resulting hierarchical construction enables scalable coarse-to-fine sampling at large lattice volumes, particularly in the critical regime where both conventional sampling methods and single-scale generative models become increasingly inefficient.

We demonstrate the method for two-dimensional scalar $\phi^4$ lattice field theory near criticality. The resulting sampler reproduces physical observables in agreement, within statistical uncertainties, with Hybrid Monte Carlo (HMC) reference runs, exhibits more favorable scaling with lattice size than existing generative baselines, and achieves orders-of-magnitude smaller autocorrelations than HMC on the largest lattices considered. Additionally, in our proposed method the coarse-to-fine hierarchy induces a built-in coupling between neighboring resolutions, providing exactly the structure needed to construct multilevel Monte Carlo (MLMC) estimators~\cite{Giles_2015,PhysRevD.102.114512} and enabling further variance reduction at fixed computational cost.

The remainder of this paper is organized as follows. In Sec.~\ref{sec:background} we review the sampling problem in continuous lattice systems near criticality. In Sec.~\ref{sec:method} we introduce the multiscale generative framework and the associated multilevel estimator. In Sec.~\ref{sec:experiments} we present numerical results for the two-dimensional scalar $ \phi^4 $ theory. Finally, Sec.~\ref{sec:conclusion} concludes with a broader discussion of possible future directions, including further developments of the multiscale generative framework and its application to more challenging lattice field theories.

\section{Background: Sampling LFT Near Criticality}
\label{sec:background}

Consider lattice field theories defined on a hypercubic lattice
$\Lambda_L = \{1,\ldots,L\}^D$ with continuous degrees of freedom
$\phi_x \in \mathbb{R}$ at each site $x \in \Lambda_L$.
A field configuration $\phi = \{\phi_x\}_{x\in\Lambda_L}$ is distributed according to
the Boltzmann measure
\begin{equation}
  p(\phi) = \frac{1}{Z}\exp\!\bigl[-S(\phi)\bigr],
  \label{eq:boltzmann}
\end{equation}
where $S(\phi)$ is a local action functional and
$Z = \int \mathcal{D}\phi\,e^{-S(\phi)}$ is the partition function.
Physical observables are expectation values
$\langle\mathcal{O}\rangle = Z^{-1}\int\mathcal{D}\phi\,\mathcal{O}(\phi)\,e^{-S(\phi)}$,
which in practice are estimated numerically from ensembles of configurations
generated by Markov chain Monte Carlo (MCMC).

A standard workhorse algorithm for continuous-variable lattice systems is
Hybrid Monte Carlo (HMC)~\cite{DUANE1987216}, which combines
molecular-dynamics trajectories with a Metropolis acceptance step to generate
global proposals for the field configuration.

\subsection{Critical Slowing Down}

A central measure of sampling efficiency is the integrated autocorrelation
time of an observable $\mathcal{O}$,
\begin{equation}
  \tau_{\rm int}[\mathcal{O}]
  = \frac{1}{2} + \sum_{t=1}^{\infty}
    \frac{C_{\mathcal{O}}(t)}{C_{\mathcal{O}}(0)},
  \label{eq:tau_int}
\end{equation}
where
$C_{\mathcal{O}}(t)
= \langle\mathcal{O}(\phi^{(0)})\mathcal{O}(\phi^{(t)})\rangle
- \langle\mathcal{O}\rangle^2$
is the autocorrelation function of the Markov chain~\cite{Madras1988,Wolff2004}.
If $M$ correlated configurations are available, the statistical error on
$\langle\mathcal{O}\rangle$ scales as
\begin{equation}
  \delta\langle\mathcal{O}\rangle \sim
  \sqrt{\frac{2\tau_{\rm int}[\mathcal{O}]}{M}}\,\sigma_{\mathcal{O}},
\end{equation}
so that the effective number of independent samples is
$M_{\rm eff} = M/(2\tau_{\rm int})$.
Achieving a fixed target precision therefore requires
$M \propto \tau_{\rm int}$ configurations, making large autocorrelation times directly expensive.

Near a second-order phase transition or near a critical point, the correlation length
$\xi \sim |g - g_c|^{-\nu}$ diverges as the coupling $g$ approaches
the critical value $g_c$.
At finite volume, $\xi$ is bounded by $L$, but the integrated autocorrelation time
typically grows with system size as a power law,
\begin{equation}
  \tau_{\rm int} \sim \xi^{z_{\rm dyn}} \sim L^{z_{\rm dyn}},
  \label{eq:csd}
\end{equation}
where $z_{\rm dyn}$ is the dynamical critical exponent of the
algorithm~\cite{wolff1990critical,Schaefer:2010hu}.
For HMC applied to scalar field theories, $z_{\rm dyn} \approx 2$~\cite{DUANE1987216},
so that the cost of obtaining $M_{\rm eff}$ independent samples scales as
\begin{equation}
  \mathcal{C}_{\rm HMC}
  \sim M_{\rm eff}\,\tau_{\rm int}\,L^D
  \sim M_{\rm eff}\,L^{D+z_{\rm dyn}}
  = M_{\rm eff}\,L^{D+2}.
  \label{eq:hmc_cost}
\end{equation}
This power-law growth becomes prohibitive near criticality at large lattice volumes,
where the long-distance physics of interest resides. It therefore constitutes a
central bottleneck for scalable lattice simulations and motivates sampling strategies
that explicitly exploit scale separation in configuration space.

\section{Multiscale Generative Sampling}
\label{sec:method}

\subsection{Renormalization of the Action}

Renormalization-group (RG) transformations provide a natural framework for describing the scale dependence of lattice field theories by systematically integrating out short-distance degrees of freedom. Consider a lattice field configuration $\phi \in \mathbb{R}^V$ distributed according to the Boltzmann measure
\begin{equation}
p(\phi) = \frac{1}{Z} \exp\!\left[-S(\phi; \{K_i^{(0)}\})\right],
\end{equation}
where $S(\phi; \{K_i^{(0)}\})$ denotes the microscopic action parametrized by a set of couplings $\{K_i^{(0)}\}$. The effective coarse-grained action $S^{(1)}$ is defined implicitly by integrating out fine degrees of freedom,
\begin{equation}
\small
\exp\!\left[-S^{(1)}(\phi^{(1)})\right]
\propto
\int \mathcal{D}\phi \;
\delta\!\left(\phi^{(1)} - B(\phi)\right)
\exp\!\left[-S(\phi; \{K_i^{(0)}\})\right],
\label{eq:rg_step}
\end{equation}
where $\phi^{(1)} = B(\phi)$ is a blocking map from the fine lattice to a coarser one. In general, the renormalized action takes the form
\begin{equation}
S^{(1)}(\phi^{(1)}) = \sum_i K_i^{(1)} \, \mathcal{O}_i(\phi^{(1)}).
\end{equation}
Iterating this construction yields a sequence of effective actions $\{S^{(\ell)}\}$ on lattices of progressively lower resolution, together with corresponding couplings $\{K_i^{(\ell)}\}$. Under RG flow, the operator basis $\{\mathcal{O}_i\}$ grows rapidly, so that the effective action generally acquires higher-order and increasingly nonlocal interactions even when the microscopic action is local. Although the exact coarse-grained action is therefore intractable in practice, the RG picture motivates a hierarchical construction in which different resolutions are modeled separately.

\begin{figure}[t]

    \centering
    \includegraphics{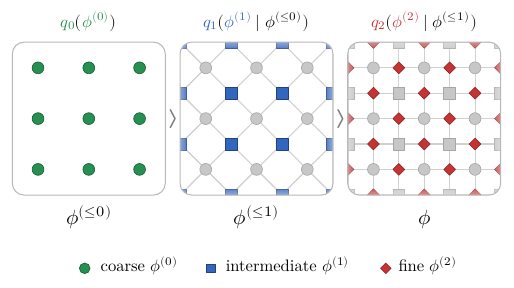}
\caption{Kadanoff-inspired multilevel partition of a periodic lattice into coarse~($\phi^{(0)}$, circles), intermediate~($\phi^{(1)}$, squares), and fine~($\phi^{(2)}$, diamonds) sites, illustrating the coarse-to-fine factorization~\eqref{eq:gen_factorization}. Each panel samples new variables (colored markers) conditioned on previously sampled sites (gray markers); connections show the nearest-neighbor dependencies used by the conditional models~$q_\ell$.}
\label{fig:multilevel}
\end{figure}

\subsection{Multiscale Generative Decomposition}

The RG picture motivates a reverse, coarse-to-fine construction in
which configurations are generated from low to high resolution.
Concretely, we introduce a multilevel partition of the lattice
sites inspired by Kadanoff blocking~\cite{Kadanoff1966} (see
\Cref{fig:multilevel}),
\begin{equation}
\phi \equiv \phi^{(\le\ell_{\max})}
\;\longrightarrow\;
\{\phi^{(0)}, \phi^{(1)}, \ldots, \phi^{(\ell_{\max})}\},
\end{equation}
where $\phi^{(0)}$ denotes the coarsest degrees of freedom and
$\phi^{(\le \ell_{\max})}$ corresponds to the full field on the finest
lattice. The exact distribution then admits the coarse-to-fine
factorization
\begin{equation}
p(\phi)
=
p(\phi^{(0)})
\prod_{\ell=1}^{\ell_{\max}}
p\!\left(\phi^{(\ell)} \mid \phi^{(\le \ell-1)}\right).
\label{eq:exact_multilevel}
\end{equation}
This decomposition would require access to the effective action at
every level. However, as discussed above, these actions are
generically non-local and not known in closed form. Working with an
approximate effective action and correcting at the finest level by a
Metropolis--Hastings step typically results in a very low acceptance
rate~\cite{Schmidt:1983rng}, which has limited the practical use of
such constructions in the traditional MCMC
setting~\cite{PhysRevD.102.114512}.

We therefore replace the unknown conditionals
$p(\phi^{(\ell)}\mid\phi^{(\le\ell-1)})$ by the learned ones and define a
hierarchical generative model
\begin{equation}
q(\phi)
=
q_0(\phi^{(0)})
\prod_{\ell=1}^{\ell_{\max}}
q_\ell\!\left(\phi^{(\ell)} \mid \phi^{(\le \ell-1)}\right),
\label{eq:gen_factorization}
\end{equation}
where $\phi^{(\le \ell)} = \{\phi^{(0)},\phi^{(1)},\ldots,\phi^{(\ell)}\}$
collects all variables up to level $\ell$. Sampling proceeds
coarse-to-fine where one first draws $\phi^{(0)}$ from $q_0$, and then
successively draws the newly introduced variables at each finer scale
from the conditional models $q_\ell$. Each $q_\ell$ is parameterized
as a flow conditioned on $\phi^{(\le\ell-1)}$, so that
$q(\phi)$ admits a tractable density and the entire model can be
trained by reverse KL minimization against the target $p(\phi)$.

A practical advantage of our decomposition is the locality. When variables at level~$\ell$ are sampled, their
nearest neighbors in the fine lattice have already been drawn at
coarser levels and are contained in $\phi^{(\le\ell-1)}$. The
conditional models $q_\ell$ therefore only need to capture short-range
correlations around the newly introduced sites, and can be implemented
with small receptive fields. This should be contrasted with a
single-scale flow acting on the full fine field, which must learn
correlations across all length scales simultaneously for which the cost
typically grows unfavorably near criticality.

Hierarchical generative constructions of related form
have recently appeared in the literature. Ref.~\cite{singha2025multilevel}
developed a learned multilevel sampler for discrete spin systems,
where coarse-to-fine conditionals are categorical and can be
parameterized directly; further variants in the discrete setting are
explored in Ref.~\cite{bialas2026hierarchical}. In the continuous case, Refs.~\cite{Abbott:2024knk,bauer2025super}
introduced multiscale flows with a different partition of the degrees
of freedom and a different training objective. Our construction
targets LFT in the regime where long-range correlations dominate
(i.e.\ near criticality), and differs from these works in both the
coarse-to-fine partition and in the form of the conditional
models~$q_\ell$, which we describe in the following subsections.

\subsection{Approximate Local Conditional Sampler with Gaussian Mixture Models}

At intermediate and finest levels (Fig~\ref{fig:multilevel}), we approximate the exact coarse-to-fine conditional distributions using local conditional models, which later be used as prior distribution for the refinement process. Concretely, we assume that the newly introduced variables at level $\ell$ are conditionally independent given the already-sampled neighboring variables in $\phi^{(\le \ell-1)}$, and model this approximation with a conditional Gaussian mixture model (GMM),
\begin{equation}
q_\ell^{\mathrm{GMM}}\!\left(z^{(\ell)} \mid \phi^{(\le \ell-1)}\right)
=
\prod_{x \in \Lambda^{(\ell)}}
q_{\ell, x}^{\mathrm{GMM}}\!\left(z_x^{(\ell)} \mid \phi_{\mathcal{NN}(x)}^{(\leq \ell-1)}\right),
\label{eq:gmm}
\end{equation}where $\Lambda^{(\ell)}$ are the lattice sites to be sampled at level $l$ and
\begin{equation}
q^{\rm GMM}_{\ell,x}\!\left(z^{(\ell)}_x \mid 
\phi^{(\leq\ell-1)}_{{\rm NN}(x)}\right)
= \sum_{k=1}^{K} \pi^{(\ell)}_{k,x} \,
\mathcal{N}\!\left(z^{(\ell)}_x \,\Big|\, 
\mu^{(\ell)}_{k,x},\, \sigma^{2\,(\ell)}_{k,x}\right).
\end{equation}
Here, the mixture weights $\pi^{(\ell)}_{k,x}$, means 
$\mu^{(\ell)}_{k,x}$, and variances $\sigma^{2\,(\ell)}_{k,x}$ 
are functions of the neighboring coarse variables 
$\phi^{(\leq\ell-1)}_{{\rm NN}(x)}$ and are parameterized 
by fully Convolutional Neural Networks.

The GMM provides an explicit, tractable density that can be sampled and evaluated efficiently, making it well suited as a local conditional prior within the hierarchical framework. It also allows parallel sampling across lattice sites.

\subsection{Conditional Sampling with Continuous Normalizing Flows}

Although, the GMM captures the leading local conditional structure at each refinement level, it does not represent the full target conditional distribution. To model the remaining mismatch, we refine GMM samples $z^{(\ell)}$ with a conditional continuous normalizing flow (CNF)~\cite{chen2018neural}. Specifically, we define an invertible transformation
\begin{equation}
\phi^{(\ell)} = f_\ell\!\left(z^{(\ell)}; \phi^{(\le \ell-1)}\right)
\end{equation}
 The map $f_\ell$ is generated by integrating a conditional neural ordinary differential equation,
\begin{equation}
\frac{d \phi(t)}{dt} = v_\ell\!\left(\phi(t), t; \phi^{(\le \ell-1)}\right),
\qquad t \in [0,1],
\end{equation}
with $\phi(0) = z^{(\ell)}$ and $\phi(1) = \phi^{(\ell)}$. The corresponding conditional density is
\begin{align}
\log q_\ell(\phi^{(\ell)} \mid \phi^{(\le \ell-1)})
&=
\log q_\ell^{\mathrm{GMM}}(z^{(\ell)} \mid \phi^{(\le \ell-1)}) \notag\\
&\quad
- \int_0^1 dt \;
\nabla \!\cdot v_\ell\!\left(\phi(t), t; \phi^{(\le \ell-1)}\right).
\end{align}
Together, the GMM and CNF provide a tractable approximation to each conditional factor in Eq.~\eqref{eq:exact_multilevel}.

\subsection{Continuous Multilevel Sampler}
The full multilevel model is obtained by combining the ingredients introduced in the previous sections across all resolutions according to Eq.~\eqref{eq:gen_factorization}.
At the coarsest level, the unconditional model $q_0$ is a CNF with a
standard Gaussian prior. At each finer level, the conditional model
$q_\ell$ is a continuous normalizing flow whose prior is the
corresponding conditional GMM. What distinguishes our construction
from existing multilevel samplers is the combination of three
ingredients: (i) our coarse-to-fine decomposition of the lattice
sites across scales, (ii) a fast, independently samplable conditional
GMM that captures the dominant local structure of the newly
introduced sites at each upsampling step, and (iii) a lightweight CNF
that learns the residual inter-site correlations on top of the GMM
prior. The full architecture, along with a comparison to related
approaches such as the Super Resolving NF (SR-NF)~\cite{bauer2025super},
is described in Appendix~\ref{app:arch}.

The model parameters of the multilevel sampler are optimized by minimizing the reverse Kullback--Leibler divergence between the hierarchical model and the target distribution,
\begin{equation}
\mathcal{L}
=
\mathrm{KL}\!\left(
q(\phi) \,\|\, p(\phi)
\right).
\end{equation}
Once trained, the model generates samples by ancestral coarse-to-fine sampling, thereby avoiding Markov-chain autocorrelations and enabling efficient exploration of configuration space near criticality.
\subsection{Multilevel Control Variates for Variance Reduction}

CNF samplers are typically used as single-level importance samplers where one generates $N$ configurations from the learned distribution $q^{(L)}$ on an $L \times L$ lattice and computes the self-normalized importance sampling (IS) estimator
\begin{equation}
  \widehat{\mathcal{O}}_{\mathrm{IS}}
  = \sum_{n=1}^{N} \bar{w}_n\,\mathcal{O}(\phi_n),
\end{equation}
\text{where}~
\begin{equation}
w(\phi) =
\frac{e^{-S(\phi)}}{q(\phi)},
\qquad 
\bar w_n =
\frac{w(\phi_n)}
{\sum_{m=1}^{N} w(\phi_m)}.
  \label{eq:is}
\end{equation}
Its variance scales as $\mathrm{Var}_{\mathrm{IS}} \propto 1/N$. The difficulty is that sampling at the finest level is expensive and each sample requires integrating all upsampling CNFs over the full $L^2$-dimensional space accordingly the per-sample cost grows roughly as $c_L \sim L^2$. Achieving a fixed target precision, therefore, becomes increasingly costly as the lattice size grows.

Multilevel Monte Carlo~\cite{Giles_2015} addresses this by exploiting the fact that expensive fine-level evaluations are only needed to correct cheaper coarse-level estimates. Since coarser levels have costs $c_\ell \ll c_L$, much more number of samples can be generated for the same computational budget.

\paragraph*{Finest-level weighted estimator.}
In practice, the target marginals $p^{(\ell)}$ are generally unavailable at intermediate levels. We therefore define importance weights only at the finest level,
\begin{equation}
w^{(L)}(\phi) =
\frac{e^{-S(\phi)}}{q(\phi)},
\qquad
\bar w^{(L)}_n =
\frac{w^{(L)}(\phi_n)}
{\sum_{m=1}^{N_L} w^{(L)}(\phi_m)}.
\end{equation}
Given samples $\{\phi^{(\le \ell)}_n\}_{n=1}^{N_\ell}$ from the hierarchical model $q^{(\ell)}(\phi^{(\le \ell)})$, and level-dependent observables $\mathcal O^{(\ell)}$ with $\mathcal O^{(L)} = \mathcal O$ and $\mathcal O^{(-1)} \equiv 0$ we define the estimator
\begin{align}
\label{mlmc}
\widehat{\mathcal O}_{\mathrm{MLMC}}
=&\;
\sum_{n=1}^{N_L}
[\bar w^{(L)}_n\,\mathcal O^{(L)}(\phi^{(\le L)}_n)
-
\frac{1}{N_L}
\mathcal O^{(L-1)}(\phi^{(\le L-1)}_n)]
\notag\\
&+
\sum_{\ell=0}^{L-1}
\frac{1}{N_\ell}
\sum_{n=1}^{N_\ell}
\Big[
\mathcal O^{(\ell)}(\phi^{(\le \ell)}_n)
-
\mathcal O^{(\ell-1)}(\phi^{(\le \ell-1)}_n)
\Big].
\end{align}

Its efficiency depends on the inter-level correlations of $\mathcal O^{(\ell)}$ and $\mathcal O^{(\ell-1)}$ reflecting the variance of each term in the form of $
\mathcal O^{(\ell)} - \mathcal O^{(\ell-1)}.$
Near criticality, the coupling between successive levels is strong, so the finer level terms have lower variances and can be accurately estimated with fewer samples. Consequently, the overall computational budget is reduced, which makes the proposed MLMC estimator especially well suited to hierarchical sampling in the critical regime.

\section{Numerical Experiments}
\label{sec:experiments}

\subsection{Model and Simulation Setup}

We study the two-dimensional scalar $\phi^4$ lattice field theory near criticality as a representative continuous lattice system with local interactions. The theory is defined on a square lattice of linear size $L$ with periodic boundary conditions where a real-valued field variable $\phi_x \in \mathbb{R}$ is assigned at each lattice site $x$.

The Euclidean lattice action is
\begin{equation}
S(\phi)
=
\sum_x \left[ (1 - 2\lambda)\,\phi_x^2 + \lambda\,\phi_x^4 \right]
- 2\kappa \sum_{x,\mu} \phi_x \phi_{x+\hat{\mu}},
\label{eq:phi4_action}
\end{equation}
where $\kappa$ denotes the nearest-neighbor coupling, $\lambda$ controls the quartic self-interaction, and the second sum runs over nearest neighbors in the two lattice directions $\mu = 1,2$. The action is invariant under the global $\mathbb{Z}_2$ symmetry $\phi_x \to -\phi_x$, which is spontaneously broken in the ordered phase. The model undergoes a second-order phase transition between symmetric and broken phases and therefore provides a standard benchmark for critical sampling in continuous lattice field theories.

We determine the critical coupling $\kappa_c$ from the Binder cumulant~\cite{Binder1981},
\begin{equation}
U_4 = 1 - \frac{\langle m^4\rangle}{3\langle m^2\rangle^2},
\qquad
m = L^{-2}\sum_x \phi_x.
\end{equation}

\begin{figure}[t]
  \centering
  \includegraphics[width=0.95\linewidth]{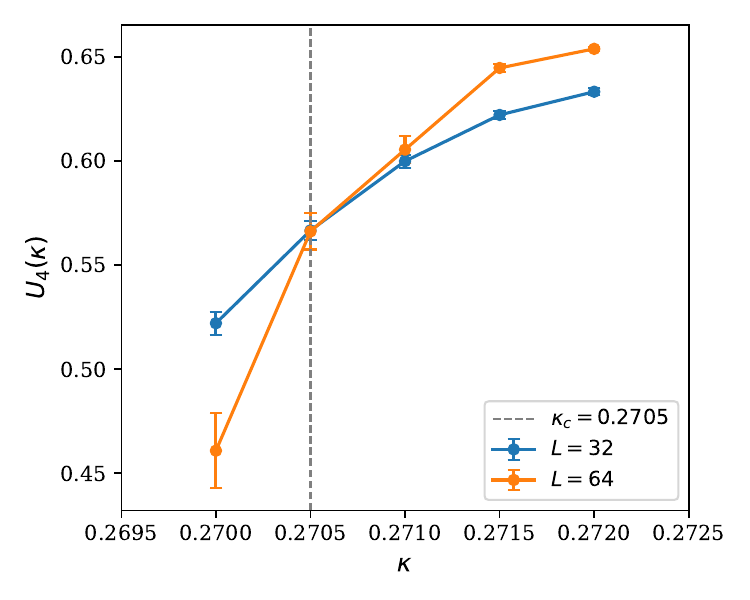}
  \caption{%
    Binder cumulant $U_4(\kappa)$ for $L=32$ and $L=64$ from HMC
    ($5\times10^5$ configurations each).
    The curves cross at $\kappa_c = 0.2705$ (dashed vertical line).
    Error bars are blocked jackknife estimates.
  }
  \label{fig:binder}
\end{figure}

As a challenging benchmark point, we choose $\kappa_c = 0.2705 \pm 0.0007$, estimated from the crossing of the Binder cumulant for $L=32$ and $L=64$. This estimate is used only to define a representative near-critical simulation point; our aim is not a precision determination of the critical coupling, but to study the method in a regime where critical slowing down is particularly pronounced.  Figure~\ref{fig:binder} shows $U_4(\kappa)$ obtained from $5\times10^5$ HMC configurations for each lattice size. The curves cross at $\kappa_c = 0.2705$, where $U_4(L=32)=0.567$ and $U_4(L=64)=0.566$ differ by only $0.001$ ($\ll 1\sigma$). This is also the regime where HMC critical slowing down is severe, with $\tau_{\rm int}(|m|)\approx{}$$4416$ at $L=128$ (Appendix~\ref{app:hmc_tau}).

We evaluate the proposed sampler near the critical point on lattices of size
$L \in \{16, 32, 64, 128, 256\}$.
As a benchmark observable, we consider the absolute magnetization
\begin{equation}
|m| = \left|L^{-2}\sum_x \phi_x\right|,
\end{equation}
and the connected susceptibility
\begin{equation}
\chi_{\rm conn}
=
L^2\left(\langle m^2\rangle - \langle |m|\rangle^2\right).
\label{eq:chi_conn}
\end{equation}

We use HMC as the reference baseline for three purposes.
First, the HMC runs provide reference estimates of physical observables, such as $\chi_{\rm conn}$ and $\langle |m| \rangle$, against which the generative samplers are tested for statistical bias. Second, they determine the autocorrelation time $\tau_{\rm int}(|m|)$, which quantifies the severity of the critical slowdown at each lattice size. Third, they define the reference computational target i.e. we measure how much wall-clock time each generative method requires to match the effective number of independent samples obtained by HMC. For $L \leq 128$ we run HMC chains of $2\times10^6$ steps; for $L=256$ we run a chain of $3\times10^6$ steps. Autocorrelation times are estimated with the Wolff method~\cite{Wolff2004,Madras1988}. Further details of the HMC simulations are given in Appendix~\ref{app:hmc_tau}.

\paragraph*{Generative methods.}
We compare four flow-based samplers, all trained at $\kappa_c = 0.2705$ with the reverse-KL objective:
\begin{itemize}
    \item \emph{Dense CNF}~\cite{gerdes2023learning}: A single CNF on the target distribution with a Gaussian prior, evaluated at $L \in \{16,32,64\}$; extending training beyond $L=64$ becomes computationally impractical (see Appendix \ref{app:dense_arch}).
    
    \item \emph{Hutch CNF}: A CNF with multiple CNN layers that uses the Hutchinson stochastic trace estimator~\cite{Hutchinson1990} to estimate the divergence stochastically, evaluated at $L \in \{16,32\}$; scaling to larger lattices becomes increasingly difficult (see Appendix \ref{app:hutch_arch}).
  
  \item \emph{Super Resolving CNF (SR-NF)}~\cite{bauer2025super}: A
hierarchical sampler based on a \emph{rectangular-flow} construction~\cite{bauer2025super} with coarsest level HMC sampling replaced by a plain CNF sampler.
At each super-resolution stage, a coarse $L_\ell\times L_\ell$ field
is first deterministically block-replicated to a fine
$2L_\ell\times 2L_\ell$ lattice, then augmented with zero-sum
Gaussian noise that injects the extra degrees of freedom, and finally
evolved through a full-lattice CNF that transports all sites jointly.
 We evaluated SR-NF at
$L\in\{16,32,64,128,256\}$ (see Appendix~\ref{app:SR-NF_arch} for more details).
\item \emph{Our Multilevel method}: The multilevel coarse-to-fine sampler introduced in Sec.~\ref{sec:method}. We evaluated our method at $L \in \{16,32,64,128,256\}$ (see Appendix \ref{app:our_arch} for more details).
\end{itemize}
\begin{table*}[t]
  \centering
  \caption{Integrated autocorrelation time $\tau_{\rm int}(|m|)$ at
    $\kappa_c=0.2705$, $\lambda=0.022$.
    HMC: Wolff method on $2\times10^6$-step chains ($L\leq128$) and
    $3\times10^6$ steps ($L=256$).
    Generative methods: IMH chains of $5\times10^4$ steps,
    $2\times10^3$ burn-in.
    The corresponding importance-sampling efficiency $\mathrm{ESS}/N$
    for all methods is reported in Fig.~\ref{fig:ess_vs_L_main}
    (Sec~\ref{sec:IS_ESS}).}
  \label{tab:tau_int_table2}
  \begin{ruledtabular}
    \begin{tabular}{lrrrrr}
      \textrm{Method} & $L=16$ & $L=32$ & $L=64$ & $L=128$ & $L=256$ \\
      \hline
      HMC & $64.6 \pm 1.3$ & $228 \pm 9$ & $1370 \pm 120$ & $4416 \pm 928$ & $8410 \pm 1991$ \\
      Our method & $0.76\pm 0.01$ & {$0.85 \pm 0.01$}& {$1.19\pm 0.01$} & {$3.69 \pm 0.06$} & {$72.2 \pm 5.5$} \\
SR-NF        & $0.89 \pm 0.01$ & $1.50 \pm 0.02$ & $5.32 \pm 0.11$ & $186 \pm 23$ & $3615 \pm 1944$ \\

      Dense CNF  & $0.70 \pm 0.01$ & $0.76 \pm 0.01$ & $0.98 \pm 0.01$ & ---             & --- \\
      Hutch CNF  & $25.7 \pm 1.2$  & $55.4 \pm 3.7$  & ---             & ---             & --- \\
    \end{tabular}
  \end{ruledtabular}
\end{table*}

For each model, physical observables are estimated from $N = 10^5$ samples using self-normalized importance sampling (IS) Eq.~\eqref{eq:is}.
Autocorrelation times for the generative samplers are measured from independent Metropolis--Hastings (IMH) chains of $5\times10^4$ steps after a burn-in of $2\times10^3$ steps, using the Sokal windowing estimator~\cite{Wolff:2003sm}. Timing measurements for all generative methods are reported as wall-clock time on the same NVIDIA A100 GPU, while the HMC timings are measured on the CPU hardware used for the reference simulations.

\subsection{Mixing and Autocorrelation Time}
\label{sec:results_tau}

Table~\ref{tab:tau_int_table2} reports the integrated autocorrelation time
$\tau_{\rm int}(|m|)$ for all methods across lattice sizes.
For HMC, the dynamical exponent $z = 1.81 \pm 0.14$ is obtained from a
power-law fit to all six measured points $L \in \{8, 16, 32, 64, 128, 256\}$
(see Appendix~\ref{app:hmc_tau}).

Three regimes can be distinguished. At small volumes ($L \leq 32$),
our method, Dense CNF, and SR-NF all achieve $\tau_{\rm int}<3$,
indicating very low autocorrelation compared with the HMC values $\tau_{int}$ 
$=64.6 \pm 1.3$ and $228\pm9$. Hutch CNF is already substantially worse
($\tau_{\rm int}=25.7 \pm 1.2$ at $L=16$), reflecting the additional variance
introduced during training by stochastic log-determinant estimation.

At $L=64$, the methods begin to separate. Our method maintains
$\tau_{\rm int}=1.19 \pm 0.01$, Dense CNF remains competitive at $0.98\pm 0.01$, whereas
SR-NF rises to $5.32\pm0.11$, signaling the onset of degraded mixing as the
full-lattice CNF struggles to represent the growing configuration space.
The contrast becomes much stronger at larger volumes where at $L = 128$ and $L = 256$, our method yields $\tau_{\rm int} = 3.69\pm0.06$ and $72.2\pm 5.5$, respectively, while SR-NF yields $186\pm 23$ at $L = 128$ and $3615\pm 1944$ at $L = 256$ (the latter has large uncertainty as shown in Table \ref{tab:tau_int_table2}, note the SR-NF's IMH acceptance rate is $\sim 0.05\%$ at $L = 256$, yielding only $\approx 7$ effective samples)

Even at $L=256$, our method remains roughly $116\times$ below the
measured HMC value of $8410 \pm 1991$ demonstrating that the
hierarchical construction substantially mitigates critical slowing down
at lattice sizes where conventional HMC becomes prohibitively expensive.
Low autocorrelation is meaningful only when accompanied by sufficient
importance-sampling efficiency; the $\mathrm{ESS}/N$ values reported
in Fig.~\ref{fig:ess_vs_L_main} confirm that our method maintains
$\mathrm{ESS}/N \geq 0.20$ up to $L=128$, where SR-NF drops to
$\mathrm{ESS}/N \approx 4\times10^{-3}$ at the same scale, rendering its low
$\tau_{\rm int}$ practically unusable at large volumes.

\subsection{Physical Observable Consistency}
\label{sec:results_obs}

\begin{figure}[t]
  \centering
  \includegraphics[width=0.95\linewidth]{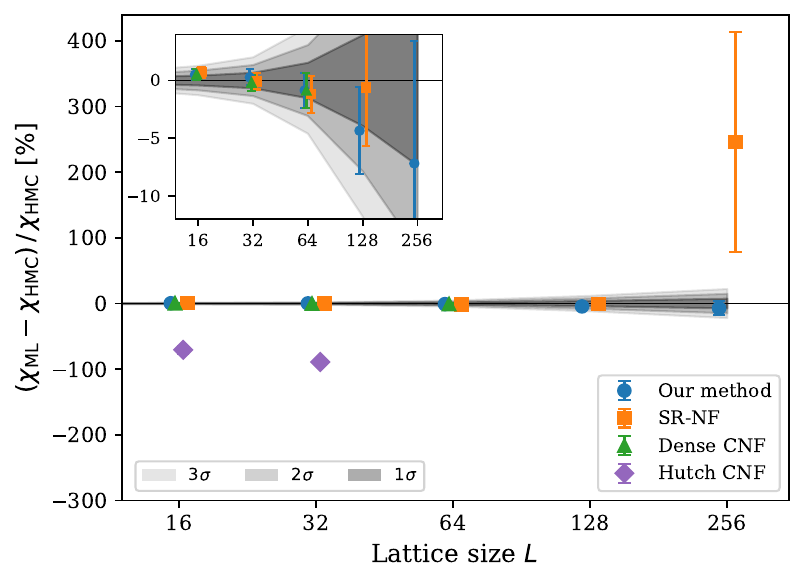}
\caption{%
Relative deviation of the connected susceptibility from HMC,
$(\chi_{\rm ML}-\chi_{\rm HMC})/\chi_{\rm HMC}$, versus lattice size $L$
at $\kappa_c=0.2705$. Grey bands show the $1\sigma$, $2\sigma$, and $3\sigma$
HMC uncertainty intervals for visual guidance. For clearer visualization of the error bars, we also included a zoomed-in view.
}
  \label{fig:chi_conn}
\end{figure}

A generative sampler is only useful if its importance-weighted estimates
remain statistically consistent with the target distribution. We test
this by comparing importance-sampling estimates of the connected
susceptibility $\chi_{\rm conn}$, obtained from $N=10^5$ samples, with
the long HMC reference runs. Figure~\ref{fig:chi_conn} shows the relative deviation
$(\chi_{\rm ML}-\chi_{\rm HMC})/\chi_{\rm HMC}$, with gray bands
indicating the uncertainty $1/2/3\,\sigma$ of the HMC reference.

The Dense CNF \cite{gerdes2023learning} and our method reproduce the HMC reference values for $\chi_{\rm conn}$ within $1\text{--}2\sigma$ across all tested lattice sizes ($L = 16$ to $256$), indicating no statistically significant bias at our level of precision. By contrast, SR-NF estimates become increasingly unreliable in large volumes: at $L = 256$, where IS efficiency drops below $0.004\%$, the reweighted $\chi_{\rm conn}$ acquires uncertainties that exceed the central value, and the mean deviates well beyond the HMC error bands. The Hutchinson-trace CNF exhibits substantial mode-dropping already at both $L = 16,32$, rendering its IS estimates systematically biased. Our method, in contrast, maintains a sub-percent relative deviation from HMC up to $L = 256$, providing stable and accurate observable estimates at scales where the baseline methods become unreliable or non-trainable. 



We also compare the histograms of the signed magnetization $m$, the absolute
magnetization $|m|$, the action density $S/V$, and the connected two-point function
of the time-slice field
$C(t)=\langle s(0)s(t)\rangle-\langle s\rangle^2$, $s(x_1)=L^{-1}\sum_{x_2}\phi_{x_1,x_2}$, with HMC simulations.

\begin{figure}[t]
  \centering
  \includegraphics[width=\columnwidth]{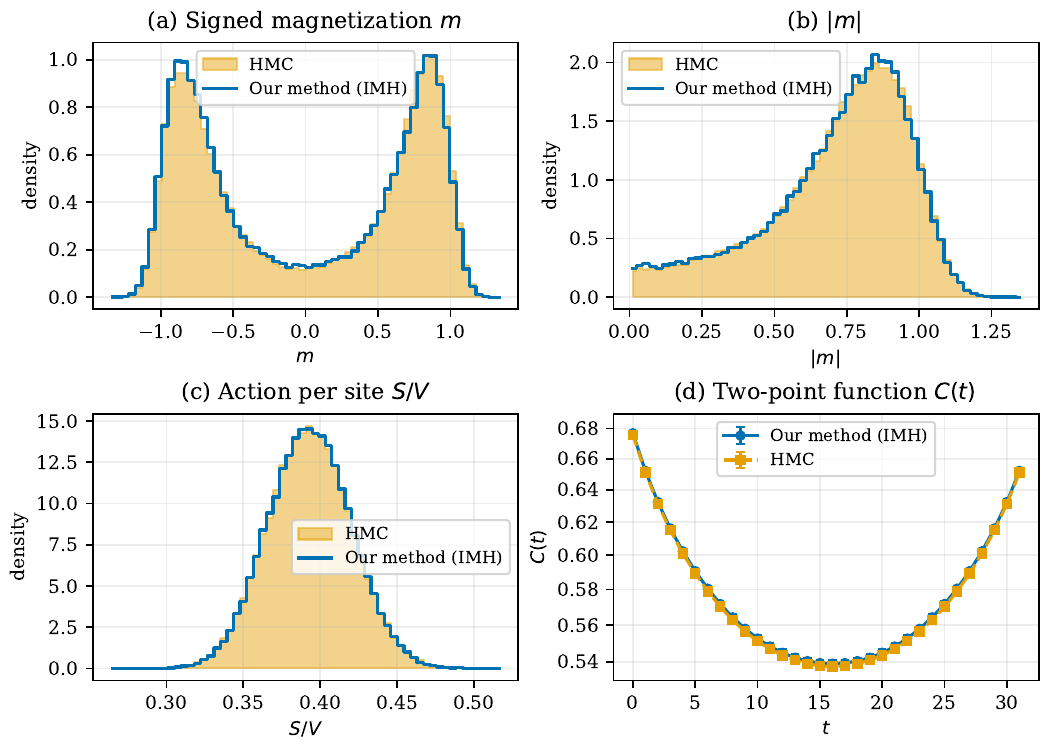}
  \caption{Observables at the finest lattice ($L=32$, $\kappa_c$) by
    our method (blue) and HMC (orange):
    (a)~signed magnetization $m$, (b)~absolute magnetization $|m|$,  (c) action density $S/V$  and (d)~the connected two-point function
    $C(t)$ of the time-slice field computed from configurations
    drawn from the model by independent Metropolis--Hastings (acceptance $\approx0.76$).}
  \label{fig:obs}
\end{figure}

For computing these observable, we do independent Metropolis--Hastings with our model as a proposal (acceptance $\approx0.76$). Figure~\ref{fig:obs} shows that the distributions of our method overlap the HMC ones for all
four observables (e.g.\ $\langle S/V\rangle={0.3931\pm0.0001}$ vs.\ $0.3931\pm0.0004$ and
$\langle|m|\rangle={0.7156\pm0.0009}$ vs.\ $0.7173\pm0.0066$), and that the two-point functions agree
within errors across the accessible range of $t$.

\subsection{IS Efficiency Comparison: ESS}
\label{sec:IS_ESS}

\begin{figure}[h]
  \centering
  \includegraphics[width=0.9\linewidth]{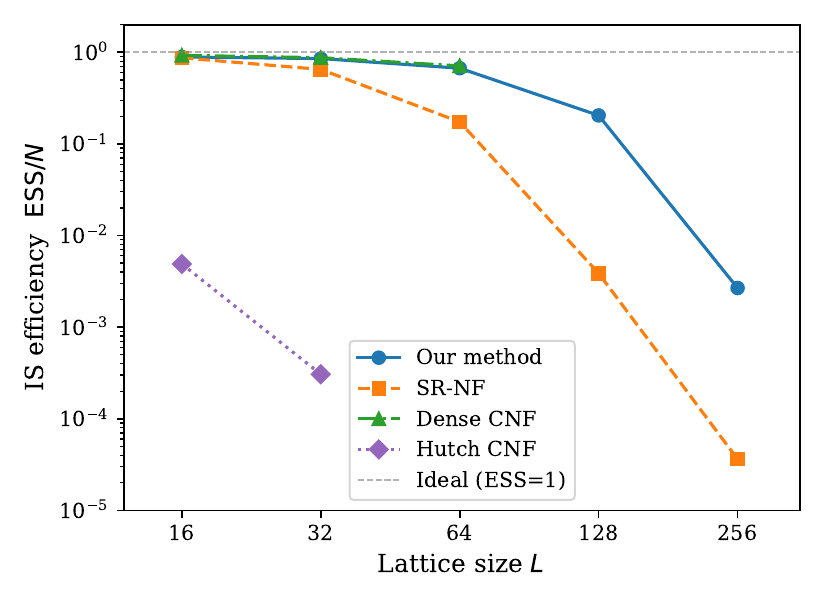}
  \caption{%
    IS efficiency $\mathrm{ESS}/N$ vs.\ $L$ at $\kappa_c=0.2705$.
    Evaluations use $N=10^5$ IS samples per method.
    Our method (blue circles) maintains high efficiency across all
    evaluated sizes; SR-NF (orange squares) collapses at large $L$;
    Dense CNF (green triangles) is limited to $L\leq 64$;
    Hutch CNF (purple diamonds) suffers from stochastic-estimator noise.
  }
  \label{fig:ess_vs_L_main}
\end{figure}

ESS per sample is computed 
as

$$\text{ESS}/N = \frac{1}{N \sum_{i=1}^{N} \bar{w}_i^2},$$
where $ \bar{w}_i$ is defined in Eq.~\eqref{eq:is}.
Figure~\ref{fig:ess_vs_L_main} shows the IS efficiency $\mathrm{ESS}/N$
at $N=10^5$ samples for all four methods.
Our method maintains $\mathrm{ESS}/N\geq 66$\% up to $L=64$ and
$24\%$ and $0.8\%$ at $L=128,256$ respectively, while SR-NF drops to $0.4\%$ at $L=128$
and collapses to $0.04\%$ at $L=256$.
Dense CNF is competitive up to $L=64$ ($\mathrm{ESS}/N=0.71$) but
cannot be trained at larger volumes.
Hutch CNF achieves near-zero efficiency even at small $L$ ($0.5\%$
at $L=16$) reflecting noisier training signal from the
stochastic log-determinant.

\subsection{Time to Match HMC Precision}

As practical performance  of a sampler depends not only on sample quality but also on how quickly samples can be generated, both ESS and the sample-generation rate must be taken into account when assessing a generative sampler. To capture this trade-off, we measure the wall-clock time required by each method to produce the same effective number of independent samples as the HMC reference run.

At each lattice size $L$, the HMC reference uses $N_{\rm HMC}$ steps
($2\times10^6$ for $L\leq128$, $3\times10^6$ for $L=256$), corresponding to
\begin{equation}
  N_{\rm eff}^{\rm HMC} = \frac{N_{\rm HMC}}{2\tau_{\rm int}}
\end{equation}
effectively independent samples. For a generative method with IS efficiency
$\mathrm{ESS}/N$ and sample-generation rate $\Gamma$, the time required to accumulate the same number of effective samples is
\begin{equation}
  T = \frac{N_{\rm eff}^{\rm HMC}}{(\mathrm{ESS}/N)\,\Gamma}.
  \label{eq:time_to_hmc}
\end{equation}
The HMC wall-clock time
\begin{equation}
  T_{\rm HMC} = N_{\rm HMC}\, t_{\rm step}
\end{equation}
is measured on the same CPU hardware used for the reference chains, with $t_{\rm step}$ benchmarked from 500 HMC steps at each lattice size.

\begin{figure}[t]
  \centering
  \includegraphics[width=0.85\linewidth]{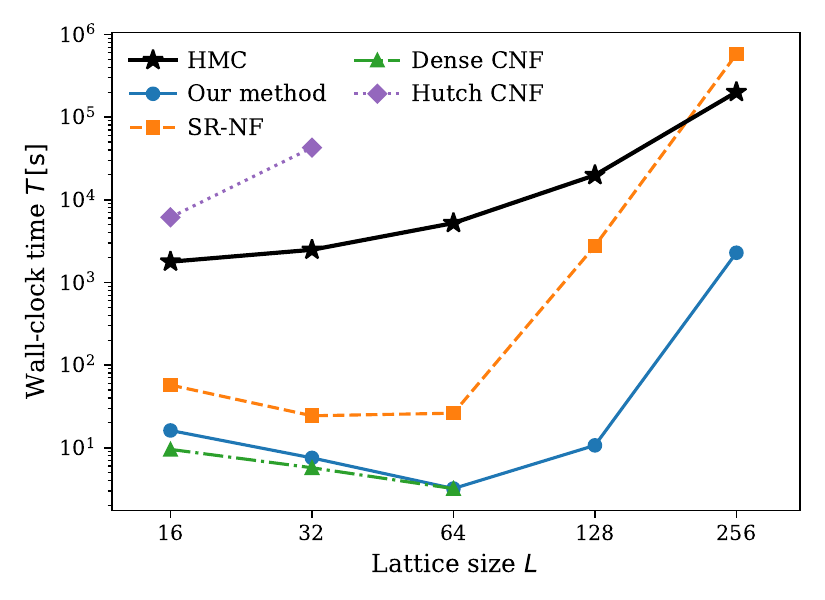}
  \caption{%
    Wall-clock time required to match the effective number of independent
    samples obtained in the HMC reference run, as a function of lattice size
    $L$ [Eq.~\eqref{eq:time_to_hmc}]. The black line shows the actual HMC
    wall-clock time.
  }
  \label{fig:time_to_hmc}
\end{figure}
Figure~\ref{fig:time_to_hmc} compares the wall clock time of generating samples with the HMC reference. Across these lattice sizes, the generative methods stay below the HMC baseline, suggesting a practical computational advantage. Our method achieves speedups 
of $106\times$ at $L=16$, $1653\times$ at $L=64$, and $211\times$ 
at $L=256$ relative to HMC. SR-NF is competitive at small $L$ 
but rises close to the HMC line at $L=128$ and $256$, where its ESS
drops to $0.4\%$ and $0.04\%$ respectively.

The inference-time comparison above excludes the one-time offline training cost. GPU training times 
for all methods and lattice sizes are reported in 
Table~\ref{tab:training_cost} (Appendix~\ref{app:training_cost}). 
At the largest lattice considered, our method requires 11.5~GPU-hours 
at $L=256$. The evolution of the training $\mathrm{ESS}/N$ as a 
function of GPU time for our method and SR-NF is shown in 
Fig.~\ref{fig:training_gpu_hours} (Appendix~\ref{app:training_efficiency}), 
demonstrating that our masked architecture reaches higher ESS 
in less training time at both $L=64$ and $L=128$. Dense CNF 
remains faster than HMC for $L\leq64$ but cannot be trained 
beyond that scale; the memory scaling underlying this limitation 
is analyzed in Appendix~\ref{app:scaling}.

\begin{figure}[t]
  \centering
  \includegraphics[width=0.92\linewidth]{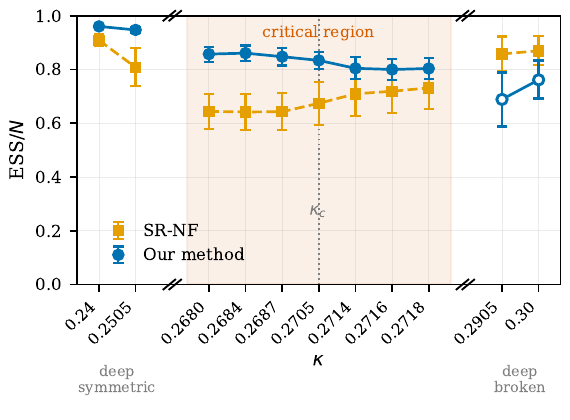}
  \caption{Importance-sampling efficiency $\mathrm{ESS}/N$ at $L=32$ across different values of $\kappa$, comparing our method to the SR-NF baseline.}
  \label{fig:kappa_ess}
\end{figure}

\subsection{Performance Across a Wider Range of Action Parameters}
\label{sec:wide_kappa}

We perform a broader scan in $\kappa$ and compare the importance-sampling efficiency of our multilevel sampler with the SR-NF baseline at $L=32$ across eleven values of $\kappa$, spanning the symmetric phase, the critical point $\kappa_c=0.2705$, and the broken phase (Fig.~\ref{fig:kappa_ess}). Seven values are chosen around the critical region, where the correlation length is largest and sampling is most challenging, while four additional values probe the deep symmetric phase ($\kappa=0.24,0.2505$) and the deep broken phase ($\kappa=0.2905,0.30$).

Our multilevel sampler achieves a higher $\mathrm{ESS}/N$ than SR-NF near criticality, and across the entire near-critical region. The improvement is most pronounced where long-range correlations are strongest; for example, $\mathrm{ESS}/N=0.84$ for our multilevel compared with $0.67$ for SR-NF at $\kappa_c$, and $0.95$ compared with $0.81$ at $\kappa=0.2505$.

At the four points farthest from criticality, both models perform well, as expected, since the correlation length is of the order of the lattice spacing. In the deep symmetric phase, at $\kappa=0.24$ and $\kappa=0.2505$, our method outperforms SR-NF, whereas in the deep broken phase, at $\kappa=0.2905$ and $\kappa=0.30$, SR-NF gives a higher $\mathrm{ESS}/N$. Nevertheless, both methods remain effective in these regimes. The main motivation for the multilevel construction, however, is to capture long-range correlations through hierarchical coarse-to-fine generation and thereby alleviate critical slowing down. This is precisely where our method shows its main advantage: in the near-critical region, where the correlation length becomes large, it consistently outperforms the SR-NF baseline. We do not include comparisons with other baseline models, as these approaches are not designed to scale systematically to larger lattice sizes.

\subsection{A qualitative picture of the performance gain}

Reproducing long-range correlations is one of the central challenges that our sampler is designed to address near criticality. Although SR-NF also follows a multilevel coarse-to-fine strategy, at each refinement level it replicates the coarse configuration onto the finer lattice, adds noise, and applies a flow that updates the entire field without further forcing the mean zero constraint. The inherited coarse values are therefore not preserved, and the long-wavelength structure may need to be transported again at every level, including on the finest lattice. In our construction, the coarse sites are instead embedded into the finer lattice and kept fixed, while masked flows generate only the newly introduced sites conditioned on their frozen coarse neighbourhood. Long-range correlations established on the coarse lattice are thus inherited directly, leaving each refinement flow to model primarily the local conditional structure or UV fluctuations.

In our sampler, each refinement level is further divided into two stages, centre sites are generated first, followed by edge sites conditioned on both the frozen coarse field and the newly generated centre sites as shown in Fig.\ref{fig:multilevel}. In contrast, the single-step upsampling used in SR-NF requires a single flow to reconstruct the full fine-lattice structure at once, which we find difficult to stabilise with a small number of ODE steps. The ablation studies in Appendix~\ref{sec:ablation} show that the conditional GMM prior plays an important role in stabilising this training. By placing the newly introduced sites close to their conditional target distributions before the flow acts, the GMM prior enables stable and efficient sampling with a relatively small ODE integration budget.

Overall, these results show that our hierarchical coarse-to-fine construction provides a substantial computational advantage near criticality when both sampling efficiency and generation cost are taken into account. 


\subsection{Multi-Level Monte Carlo Variance Reduction}
\label{sec:results_mlmc}
\begin{table}[h]
  \centering
  \caption{%
    Comparison of plain IS and MLMC estimators for $\langle |m| \rangle$
    using our method at $\kappa_c = 0.2705$, evaluated at equal
    computational budget $T$.
  }
  \label{tab:mlmc}
  \begin{ruledtabular}
  \begin{tabular}{ccccc}
    $L$ & $T$ [s] & IS estimate & MLMC estimate & Variance reduction \\
    \hline
    32 & 614  & $0.7143 \pm 0.00085$ & $0.7130 \pm 0.00067$ & $38\%$ \\
    64 & 1266 & $0.5982 \pm 0.00102$ & $0.5996 \pm 0.00091$ & $20\%$ \\
  \end{tabular}

  \end{ruledtabular}
\end{table}

A key advantage of our hierarchical construction is that it naturally enables multilevel Monte Carlo (MLMC) estimation. Now we compare the MLMC estimator defined in~\eqref{mlmc} against plain importance sampling (IS) at equal wall-clock budget,
\[
T = N_{\rm IS}\, c_L,
\]
with $N_{\rm IS}=10^5$, where $c_L$ is the measured per-sample cost at the finest level. Per-level costs $c_\ell$ and correction variances $\sigma_\ell^2$ are estimated from a probe run, and the allocation of the sample number is then chosen according to the standard MLMC rule
\begin{equation}
  N_\ell \propto \sqrt{\sigma_\ell^2 / c_\ell}.
  \label{eq:optimal_n}
\end{equation}
As an observable, we use absolute magnetization $|m|$. We compute our estimator for $L=32$ and $L=64$, with the coarsest level fixed at $L_0=2$.
 Table.~\ref{tab:mlmc} summarizes the results of our estimator. The MLMC estimator achieves lower variance than plain IS at the same computational budget at both lattice sizes. At $L=32$, the variance reduction corresponds to a $38\%$ savings in wall-clock time for a fixed precision target. At $L=64$, the speedup is $20\%$ and the lower gain at $L=64$ can be traced to the reduction in the finest-level of ESS, from $\mathrm{ESS}/N = 85\%$ at $L=32$ to $66\%$ at $L=64$. The achievable gain is currently limited by the quality of the finest-level model, indicating that further improvements in fine-level modeling could directly enhance the MLMC speedup. So, maintaining a sufficiently high ESS at larger volumes
is therefore necessary to obtain a scalable MLMC improvement. Note that in SR-NF the observables at adjacent levels do not couple as strongly as in our decomposition, and hence the MLMC estimator does not provide any computational gain over the plain IS estimator. The possibility of applying MLMC to SR-NF is discussed in Appendix~\ref{app:SR-NF_mlmc}.
 


\section{Conclusion}
\label{sec:conclusion}
We have introduced a multiscale generative sampling framework for continuous lattice field theories near criticality. The method combines a hierarchical coarse-to-fine factorization of the target distribution with learned local conditional priors and flow-based refinement across resolutions. The important long-range correlation is modeled at coarse levels, whereas finer resolutions are generated through conditional GMM priors and continuous normalizing flows that refine the newly introduced variables while preserving the embedded coarse sites.

Applied to the two-dimensional scalar theory $\phi^4$ near criticality, the proposed method shows clear advantages over all flow-based baselines considered in this work. It achieves substantially lower integrated autocorrelation times at the large lattice sizes, while maintaining higher ESS than baseline generative models. This advantage is also reflected in wall-clock efficiency, where among all generative baselines considered, our method requires the least time to match the effective precision of the HMC reference, with the benefit becoming most pronounced at large lattice sizes. Physical observables are reproduced in statistical agreement with long HMC reference simulations across all tested lattice sizes. In addition, the hierarchical construction enables the use of an MLMC estimator, which provides variance reduction near criticality beyond plain importance sampling.

Several directions remain open. The decrease in ESS at the largest volumes suggests that further gains may be possible through larger models or improved training at the finest levels, which would likely also enhance the attainable MLMC speedup. More broadly, the present results establish hierarchical generative sampling as a promising strategy for continuous lattice systems in the critical regime. An important next step is the extension of this framework to continuous group-valued theories such as lattice gauge theories. A natural extension would be to formulate the coarse-to-fine hierarchy directly in terms of plaquette, or field-strength variables rather than link variables. This would preserve the desirable feature that each lattice site carries only a small number of local degrees of freedom, allowing the same masked, level-wise refinement strategy to be applied. The main conceptual challenge is to adapt the construction to group-valued variables while consistently enforcing the associated gauge constraints; we leave this development to future work.

\begin{acknowledgments}
This work was supported by the German Ministry for Education and Research (BMBF) under the grant BIFOLD25B, and the European Union’s HORIZON MSCA Doctoral Networks programme project AQTIVATE (101072344). We thank Stefan Kühn for helpful discussions and valuable comments on the manuscript.

\end{acknowledgments}
\section*{Data Availability}

The code used to generate the results presented in this article, together
with the pretrained models and evaluation data, is available at
\url{https://git.tu-berlin.de/rigcs/phi4/multilevel_phi4}.

\bibliographystyle{apsrev4-2}
\bibliography{apssamp}
\appendix

\section{HMC Simulation: Critical Slowing Down and Scaling}
\label{app:hmc_tau}

\subsection{Simulation Details}
For $L \in \{8, 16, 32, 64, 128, 256\}$ we run HMC chains with molecular-dynamics
trajectory length $\tau_{\rm MD}=1$. For $L\le 64$ we use $n_{\rm leap}=10$ leapfrog
steps ($\epsilon=0.1$); at the largest volumes the integrator is re-tuned to maintain a
$\gtrsim 80\%$ acceptance, using $n_{\rm leap}=14$ ($\epsilon\simeq0.071$) at $L=128$ and
$n_{\rm leap}=20$ ($\epsilon=0.05$) at $L=256$, consistent with the expected
$\epsilon\propto V^{-1/4}$ scaling.
All simulations are performed at $\kappa_c = 0.2705$, $\lambda = 0.022$
(acceptance rates in Table~\ref{tab:hmc_acc}).

\begin{table}[h]
  \centering
  \caption{%
    HMC acceptance rate and wall-clock time for the reference chains
    at $\kappa_c=0.2705$, $\lambda=0.022$, $\tau_{\rm MD}=1$(
    $n_{\rm leap}$ re-tuned per volume for $L\ge128$; see text)}.
  
  \label{tab:hmc_acc}
  \begin{ruledtabular}
  \begin{tabular}{cccc}
    $L$ & Chain length & Acceptance & Wall-clock time \\
    \hline
     8  & $2\times10^6$ & $98\%$ & $0.5$\,hr \\
    16  & $2\times10^6$ & $96\%$ & $0.5$\,hr \\
    32  & $2\times10^6$ & $91\%$ & $0.7$\,hr \\
    64  & $2\times10^6$ & $85\%$ & $1.5$\,hr \\
   128  & $2\times10^6$ & $85\%$ & $5.5$\,hr \\
   256  & $3\times10^6$ & $82\%$ & $55.8$\,hr \\
  \end{tabular}
  \end{ruledtabular}
\end{table}

The integrated autocorrelation time $\tau_{\rm int}(|m|)$, shown in \Cref{tab:hmc_tau}, is estimated from
each chain using the Wolff method~\cite{Wolff2004,Madras1988}, the
Sokal automatic windowing procedure.

\subsection{Power-Law Fit}

\begin{figure}[h]
  \centering
  \includegraphics[width=0.9\linewidth]{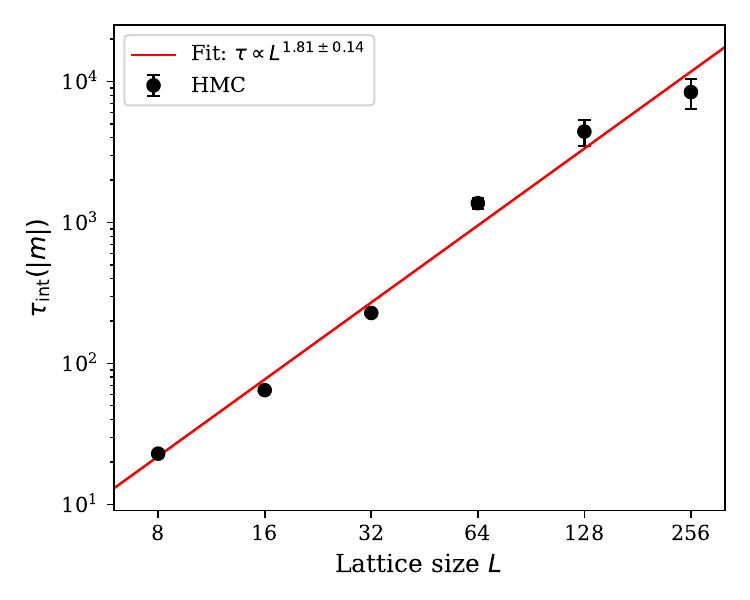}
  \caption{%
    HMC integrated autocorrelation time $\tau_{\rm int}(|m|)$ vs.\ $L$
    on a log-log scale at $\kappa_c=0.2705$.
    Filled circles: measured values (Table~\ref{tab:hmc_tau}).
    Red line: power-law fit $\tau_{\rm int} = 0.37\,L^{1.81}$
    ($z_{\rm dyn} = 1.81 \pm 0.14$), fitted to six points.
  }
  \label{fig:tau_scaling}
\end{figure}

Table~\ref{tab:hmc_tau} collects the measured $\tau_{\rm int}(|m|)$ values.
The dynamical critical exponent $z_{\rm dyn}$ is extracted by fitting the
power-law ansatz $\tau_{\rm int} = A\,L^{z_{\rm dyn}}$ in the log-log space
using a maximum likelihood fit (with the Gaussian noise model) to six measured points at
$L \in \{8,16,32,64,128,256\}$ as shown in Fig~\ref{fig:tau_scaling}:
\begin{equation}
  z_{\rm dyn} = 1.81 \pm 0.10,
  \qquad
  A = 0.14,
  \label{eq:hmc_z}
\end{equation}
where the uncertainty is a bootstrap standard deviation ($50{,}000$ resamples).
This is consistent with the theoretically expected value $z_{\rm dyn} \approx 2$
for HMC applied to scalar field theories~\cite{DUANE1987216}, and with
other measurements~\cite{wolff1990critical,PhysRevD.102.114512}.
The $L=256$ measurement ($\tau_{\rm int} = 8{,}410 \pm 1{,}991$) has
a relative uncertainty of $24\%$, reflecting the effective
sample size $N_{\rm eff} = 3\times10^6/(2\times8{,}410)\approx 178$.
\begin{table}[h]
  \centering
  \caption{%
    HMC integrated autocorrelation time $\tau_{\rm int}(|m|)$ at
    $\kappa_c=0.2705$, $\lambda=0.022$, measured by the Wolff
    method~\cite{Wolff2004}.
  }
  \label{tab:hmc_tau}
  \begin{ruledtabular}
  \begin{tabular}{cccc}
    $L$ & Chain length & $\tau_{\rm int}(|m|)$ & Note \\
    \hline
     8  & $2\times10^6$ & $22.9 \pm 0.3$       & \\
    16  & $2\times10^6$ & $64.6 \pm 1.3$       & \\
    32  & $2\times10^6$ & $228 \pm 9$          & $\tau_{\rm int} \propto L^{1.81\pm0.14}$ \\
    64  & $2\times10^6$ & $1370 \pm 120$       & \\
   128  & $2\times10^6$ & $4416 \pm 928$       & \\
   256  & $3\times10^6$ & $8{,}410 \pm 1991$  & $N_{\rm eff}\approx 178$ \\
  \end{tabular}
  \end{ruledtabular}
\end{table}

\section{Model Architecture Details and Hyperparameters}
\label{app:arch}

\subsection{Our Multilevel Method}
\label{app:our_arch}

\paragraph{Overall structure.}
Our method is a hierarchical coarse-to-fine sampler. It starts from a
coarse $8\times8$ lattice and applies $K=\log_2(L/8)$ successive
upsampling steps, each doubling the linear lattice size. The full
model therefore consists of a standalone coarse-level CNF followed by
$K$ upsampling modules, where each upsampling module is composed of two
conditional Gaussian mixture model (GMM) and two masked convolutional
CNFs.

\paragraph{Coarse-level CNF.}
The coarse $8\times8$ field is generated by a plain dense CNF, whose
architecture follows~\cite{gerdes2023learning} with different
hyperparameters. It uses an FFT-based vector field, integrated with the
RK4 solver for $n_{\rm steps}^{\rm coarse}=40$ function evaluations (when considered as baseline, in the dense CNF we have used 50 steps).
The vector field is conditioned on a Fourier basis with $n_{\rm sin}=15$
sine harmonics and $n_{\rm time}=8$ time-kernel features; bond
(nearest-neighbour) interactions are encoded with
$n_{\rm freq}^{\rm bond}=18$ and $n_{\rm time}^{\rm bond}=18$. Although
this coarse CNF acts on only $64$ sites, it is expressive enough to
capture the long-wavelength modes, the dominant source of
critical slowing down as the lattice at this scale is small
enough for a single dense flow to learn global correlations directly.

\paragraph{Upsampling step.}
When refining an $L_\ell\times L_\ell$ lattice to
$2L_\ell\times2L_\ell$, the fine lattice is partitioned into
even-even coarse sites (which carry the already-sampled coarse field),
odd-odd body-centre sites, and edge sites $(2i+1,2j)$ and $(2i,2j+1)$.
The even-even sites remain fixed throughout the upsampling step, while
the odd-odd and edge sites are generated sequentially (see
Fig.~\ref{fig:multilevel}):
\begin{enumerate}
  \item \emph{Body-centre generation.} A conditional GMM produces a
    local proposal for each body-centre site, conditioned on its
    nearest coarse neighbours. A masked convolutional CNF then refines
    the full fine lattice, with its vector field masked to zero on
    the even-even and edge sites so that only body-centre values are
    modified by the ODE flow.
  \item \emph{Edge generation.} The same procedure is applied to the
    edge sites, conditioned on the now-complete coarse and body-centre
    variables. Again, the CNF vector field is masked to zero everywhere
    except on edge sites.
\end{enumerate}
Because the CNF velocity field vanishes on the previously sampled
coordinates, those values are preserved exactly under the ODE
integration, and the restriction $\phi^{(\ell+1)}\to\phi^{(\ell)}$
obtained by reading off the even-even positions recovers the coarse
sample without error. This exact restriction is what enables the
MLMC estimator in Sec.~\ref{sec:results_mlmc}.

\paragraph{Role of the GMM prior.}
At each upsampling level, the GMM provides a conditional prior for the
newly introduced variables. It is fully factorized across the new sites,
i.e.\ the sites $\phi^{(\ell)}$ are \emph{conditionally independent} given
the already-sampled coarse field $\phi^{(\leq\ell-1)}$. Each site
distribution is a $K$-component mixture of Gaussians whose parameters
$(\pi_{k,x}^{(\ell)},\mu_{k,x}^{(\ell)},\sigma_{k,x}^{2(\ell)})$ are
output by a small convolutional network that sees only a local patch
of the coarse field around $x$. This design has two benefits.
First, it concentrates the bulk of the short-distance variability into
a cheap, tractable, closed-form density that can be sampled and
evaluated in parallel across all sites. Second, it supplies the
subsequent CNF with a sample already close to the correct local
distribution, so the CNF only needs to learn the \emph{residual}
inter-site correlations.

\paragraph{Gradient estimation for the conditional GMM prior}
\label{app:gmm_grad}

At each refinement level the newly introduced sites $z$ are drawn from a
conditional Gaussian mixture prior
\begin{equation}
  q_\theta(z\mid c) = \sum_{k=1}^{K} \pi_k(c;\theta)\,
  \mathcal{N}\!\big(z;\,\mu_k(c;\theta),\,\sigma_k^{2}(c;\theta)\big),
  \label{eq:gmm_prior}
\end{equation}
whose weights $\pi_k$, means $\mu_k$, and standard deviations $\sigma_k$ are
produced by a shared network conditioned on the coarse field $c$.

Sampling proceeds in two steps. A component index is first drawn from the
categorical weights, $k\sim\mathrm{Cat}(\pi_1,\dots,\pi_K)$, via a hard
inverse-CDF draw. Conditional on $k$, the value is generated by
reparameterization,
\begin{equation}
  z = \mu_k(c;\theta) + \sigma_k(c;\theta)\,\varepsilon,
  \qquad \varepsilon\sim\mathcal{N}(0,\mathbf{I}),
  \label{eq:gmm_reparam}
\end{equation}
so that pathwise gradients reach $\mu_k$ and $\sigma_k$.

The model is trained by minimizing the reverse Kullback--Leibler objective
$\mathcal{L}(\theta)=\mathbb{E}_{\phi\sim q_\theta}[\,\ell_\theta(\phi)\,]$ with
$\ell_\theta(\phi)=\log q_\theta(\phi)+S(\phi)$ (up to an additive constant).
Because the sampling distribution depends on $\theta$ through both the
continuous parameters and the discrete component index, its gradient contains
two contributions,
\begin{equation}
  \nabla_\theta \mathcal{L}
  = \underbrace{\mathbb{E}\big[\nabla_\theta \ell_\theta(\phi)\big]}
    _{\text{pathwise}}
  + \underbrace{\mathbb{E}\Big[\big(\bar{\ell}_\theta(\phi)-b\big)
    \!\!\sum_{\text{levels}}\!\nabla_\theta\log \pi_{k}(c;\theta)\Big]}
    _{\text{score function (weights)}},
\end{equation}
where $\bar{\ell}_\theta=\mathrm{sg}[\log q_\theta(\phi)+S(\phi)]$ is the
per-sample objective used as a detached reward, $\mathrm{sg}[\cdot]$ denotes
stop-gradient, and $b=\langle\bar{\ell}_\theta\rangle$ is a batch-mean baseline.
We use unit weight on the score term, so that Eq.~(B3) is precisely the gradient
$\nabla_\theta\mathcal{L}$ of the reverse-KL objective; the baseline $b$ is a
zero-mean control variate ($\mathbb{E}[\,b\,\nabla_\theta\log\pi_k]=0$) that
reduces its variance without biasing the estimator.

To limit the variance introduced by the score-function term, we apply it only
to the mixture-weight readout head: the logits are computed from a detached copy
of the shared trunk features, $\eta = \mathrm{head}(\mathrm{sg}[h])$ with
$\pi = \mathrm{softmax}(\eta)$, so the REINFORCE gradient updates the weight head
without perturbing the shared representation used by the means, standard
deviations, and the downstream flow.

\paragraph{Why small convolutional CNFs still scale.}
The upsampling CNFs use relatively shallow convolutional vector fields
with small kernels (Table~\ref{tab:arch_params}) and only
$n_{\rm steps}=20$ ODE steps. Individually, such networks have limited
receptive field and cannot directly represent long-range correlations
on the fine lattice. Scalability is recovered by the hierarchical
decomposition where long-wavelength structure is already encoded in the
coarse field by the coarse-level CNF, and each subsequent upsampling
level only needs to introduce short-distance fluctuations on top of it.

\paragraph{GMM and CNF hyperparameters.}
The GMM network at each upsampling level uses two convolutional layers,
$32$ hidden channels, and kernel size $5\times5$, with $H=6$ mixture
components at intermediate levels and $H=12$ at the finest level.
The body-centre and edge CNFs use convolutional vector fields with
periodic boundary conditions, the same Fourier-basis conditioning as
the coarse CNF, and RK4 integration with $n_{\rm steps}=20$. Kernel
sizes follow the schedule in Table~\ref{tab:arch_params}.
\subsection{SR-NF}
\label{app:SR-NF_arch}

SR-NF follows the coarse-to-fine upsampling architecture
of~\cite{bauer2025super}, with the HMC-based coarsest sampler of
the original proposal replaced by the same dense CNF used in our
method (RK4, 40 steps). At each upsampling step, the coarse field is
naively block-replicated on a $2\times2$ grid, Gaussian block noise is
injected, and a full-lattice (unmasked) CNF refines all
$L_{\ell+1}^2$ sites jointly.

The hyperparameter choices for SR-NF are dictated by a sequence of
coupled constraints. First, with the coarse lattice set at
$L_0 = 8$ (to match our method), we observed that SR-NF training
became unstable: the ESS initially rose, then collapsed as the loss
diverged irreversibly (see Appendix~\ref{app:scaling} for details).
Training stability was recovered only by pushing the coarse lattice
down to $L_0 = 2$, which replaces each large-stride upsampling step
with a deeper chain of smaller $2\times2$ refinements
($K = \log_2(L/2)$ levels, compared to $\log_2(L/8)$ for our method).
Second, this deeper hierarchy increases peak training memory
substantially; to fit on an A100 80\,GB GPU at $L = 128,256$ we had to
reduce the convolutional kernel size (see Table~\ref{tab:arch_params}).
Because the receptive field compounds across the extra levels, the
\emph{effective} receptive field on the finest lattice remains
comparable to that of our method despite the smaller per-level kernels.
Finally, the number of ODE steps is $n_{\rm steps} = 70$ at all sizes
except $L = 256$, where $n_{\rm steps} = 60$ was needed for  memory constraint.

\subsection{Dense CNF}
\label{app:dense_arch}

Dense CNF \cite{gerdes2023learning} is a single-scale baseline: one CNF acts directly on the
full $L\times L$ lattice, starting from an isotropic Gaussian prior.
The vector field is a \emph{dense} single-layer FFT-based
parameterisation in which each spatial mode interacts with all others,
with $\mathcal{O}(L^4)$ parameters.  It is integrated with RK4
for $n_{\rm steps}=50$ evaluations.  A richer Fourier basis is used:
$n_{\rm sin}=49$, $n_{\rm time}=11$, $n_{\rm freq}^{\rm bond}=20$,
$n_{\rm time}^{\rm bond}=20$.  The $\mathcal{O}(L^4)$ memory cost of
the exact log-determinant makes training infeasible beyond $L=64$.
Dense CNF was run at $L\in\{16,32,64\}$ with batch size $256,128,64$ and $3,000,5,000,10,000$ epochs.

\subsection{Hutch CNF}
\label{app:hutch_arch}
Hutch CNF is also a single-scale baseline acting directly on the full lattice.
Its vector field is a 3-layer convolutional network with 32 hidden channels,
$7\times 7$ kernels, and GELU activations. The divergence is estimated with
the Hutchinson stochastic trace. The ODE is solved with a fixed-step RK4 integrator at
$n_{\mathrm{steps}}=40$, and gradient checkpointing is used to reduce memory.
The same Fourier
basis as our method is used ($n_{\rm sin}=15$, $n_{\rm time}=8$).
Hutch CNF was run at $L\in\{16,32\}$ with batch size $256,128$ and
$3{,}000,5{,}000$ epochs.

\subsection{Training}
\label{app:training}

All methods are trained end-to-end with the reverse-KL objective
using the Adam optimizer~\cite{kingma2017adammethodstochasticoptimization} ($\beta_1=0.8$, $\beta_2=0.9$).  Our method
and SR-NF use initial learning rate $10^{-2}$; Dense CNF uses
$5\times10^{-3}$ as in Ref.~\cite{gerdes2023learning}.  All methods apply an exponential learning-rate
decay with $\gamma=0.9994$ and minimum $10^{-4}$.  Gradient norms are
clipped at $0.5$ for our method, SR-NF, and Hutch CNF, and at
$1.0$ for Dense CNF.  All runs use a single NVIDIA A100 GPU.
\begin{table*}[t]
  \centering
  \caption{%
    Per-size hyperparameters at $\kappa_c=0.2705$.
    Our method uses $L_0=8$; SR-NF uses $L_0=2$.
    $K$ denotes the number of upsampling levels.
    $n_{\rm steps}$ is the fine-level ODE budget per upsampling CNF;
    the coarse dense CNF always uses 40 steps.
  }
  \label{tab:arch_params}
  \begin{ruledtabular}
  \begin{tabular}{c c l c c c l c c}
    & \multicolumn{4}{c}{Our method} & \multicolumn{4}{c}{SR-NF} \\
    \hline
    $L$ & $K$ & Kernel sizes & $n_{\rm steps}$ & Batch & $K$ & Kernel sizes & $n_{\rm steps}$ & Batch \\
    \hline
    16  & 1 & $[7]$                  & 20 & 256 & 3 & $[9,\,9,\,7]$                  & 70 & 200 \\
    32  & 2 & $[9,\,7]$              & 20 & 128 & 4 & $[9,\,9,\,9,\,7]$              & 70 & 128 \\
    64  & 3 & $[9,\,9,\,7]$          & 20 & 100 & 5 & $[9,\,9,\,9,\,9,\,7]$          & 70 & 100 \\
    128 & 4 & $[9,\,9,\,7,\,7]$      & 20 &  80 & 6 & $[7,\,7,\,7,\,7,\,7,\,7]$      & 70 &  32 \\
    256 & 5 & $[9,\,9,\,7,\,7,\,7]$  & 20 &  22 & 7 & $[5,\,5,\,5,\,5,\,5,\,5,\,5]$  & 60 &  12 \\
  \end{tabular}
  \end{ruledtabular}
\end{table*}

\subsection{Hyperparameters}

Table~\ref{tab:arch_params} lists hyperparameters that vary with
target lattice size $L$ for the two hierarchical methods.

\subsection{Training and Sampling Time}
\label{app:training_cost}

Table~\ref{tab:training_cost} reports the total GPU time consumed
during training for each method and lattice size, all on a single
NVIDIA A100 80\,GB GPU.  Our method and SR-NF scale moderately
with $L$, reaching $11.5$ and $23.3$ GPU-hours at $L=256$,
respectively.  Hutch CNF is disproportionately expensive at $L=32$
($4.1$ GPU-hours) relative to its model capacity, reflecting the
high memory and compute overhead of the
graph-retention divergence estimator discussed in
Appendix~\ref{app:scaling}.
Dense CNF is cheap at small $L$ but cannot be trained beyond $L=64$.

\begin{table}[h]
  \centering
  \caption{%
    Training cost in GPU-hours (NVIDIA A100 80\,GB) at
    $\kappa_c=0.2705$, $\lambda=0.022$.
    Dashes indicate the lattice sizes for which training is infeasible.
  }
  \label{tab:training_cost}
  \begin{ruledtabular}
  \begin{tabular}{ccccc}
    $L$ & Our method & SR-NF & Dense CNF & Hutch CNF \\
    \hline
    16  & 0.3  & 0.8  & 0.3  & 0.6  \\
    32  & 0.8  & 2.2  & 0.5  & 4.1  \\
    64  & 2.3  & 5.3  & 1.5  & ---  \\
    128 & 7.8  & 18.3 & ---  & ---  \\
    256 & 11.5 & 23.3 & ---  & ---  \\
  \end{tabular}
  \end{ruledtabular}
\end{table}

Table~\ref{tab:throughput} reports the raw sampling throughput
(generated samples per second) for each method at $\kappa_c=0.2705$,
measured during the importance-sampling evaluation on
an NVIDIA A100 80\,GB GPU.

\begin{table}[h]
  \centering
  \caption{%
    Raw sampling throughput [samples/s] at $\kappa_c=0.2705$,
    $\lambda=0.022$.  Measured during IS evaluation with $N=100\,000$
    samples (batch size 256 for $L\leq64$, 128 for $L=128$, 64 for $L=256$).
    Dashes indicate the lattice sizes for which training is infeasible.
  }
  \label{tab:throughput}
  \begin{ruledtabular}
  \begin{tabular}{ccccc}
    $L$ & Our method & SR-NF & Dense CNF & Hutch CNF \\
    \hline
    16  & 995  & 310  & 1750 & 516 \\
    32  & 677  & 277  & 877  & 336 \\
    64  & 337  & 160  & 321  & ---  \\
    128 & 103  &  21  & ---  & ---  \\
    256 &  30  &   9  & ---  & ---  \\
  \end{tabular}
  \end{ruledtabular}
\end{table}


\begin{figure*}[t]
\centering
\includegraphics[width=\textwidth]{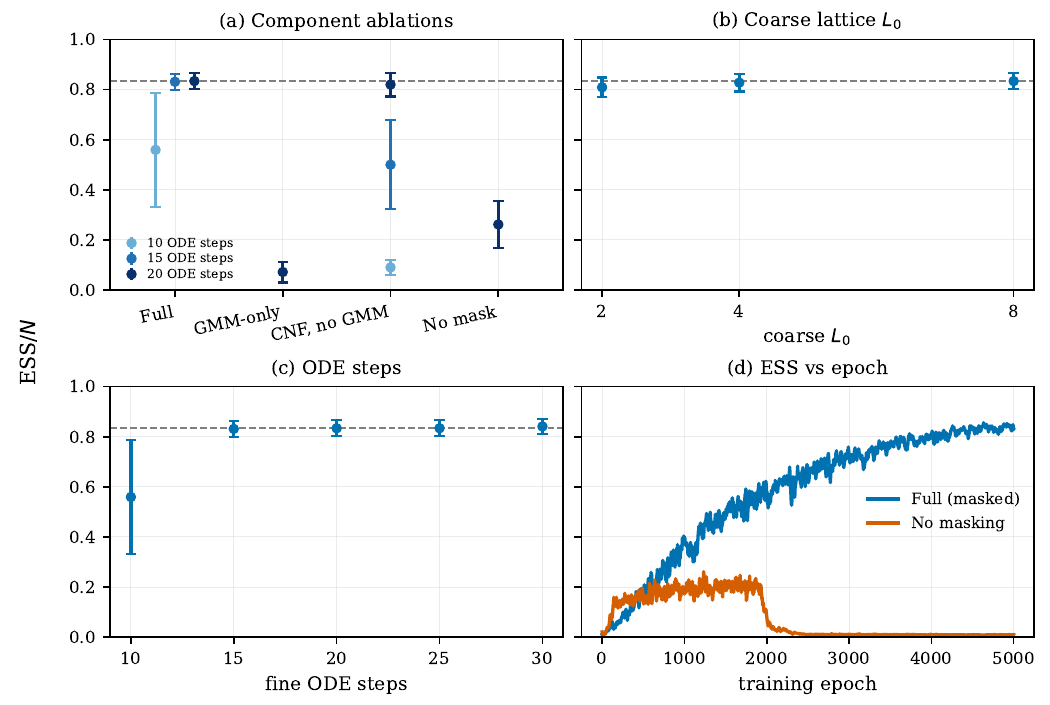}
\caption{Ablation studies at $\kappa_c=0.2705$ on an $L=32$ lattice. The reported efficiency is $\mathrm{ESS}/N$, and the error bars show the per-epoch variation of the ESS over the converged training window.
(a)~Removing individual components shows that both the masked CNF refinement and the masking structure are essential and GMM plays an important role in cost reduction by achieving high ESS at lowered numbers of ode steps.
(b)~The dependence on the initial coarse lattice $L_0$ is weak.
(c)~The efficiency saturates at approximately $15$ fine ODE steps per flow.
(d)~The unmasked full-lattice model is unstable, its ESS initially rises and then collapses, whereas the masked model converges stably.}
\label{fig:ablation}
\end{figure*}

\section{Ablation study}
\label{sec:ablation}

We evaluate the main architectural and numerical components of the sampler at the critical point $\kappa_c=0.2705$ on an $L=32$ lattice. Unless stated otherwise, all hyperparameters are kept fixed. The resulting importance-sampling efficiencies, $\mathrm{ESS}/N$, are reported in Fig.~\ref{fig:ablation} and Table~\ref{tab:gmm_ode}. We examine five components.

\emph{Full model.}
We evaluate the ESS of the full model using different numbers of ODE integration steps in the CNF. The ESS remains sufficiently higher $56\%$ even with as few as $10$ ODE steps, far above the value obtained without
the conditional prior $0.09$ indicating that the combined model is numerically stable and robust to a relatively coarse ODE discretization.

\emph{Flow refinement: GMM only}
Removing the masked flow and sampling the newly introduced sites solely from the conditional Gaussian mixture reduces the ESS from $0.83$ to $0.09$. This demonstrates that the flow refinement is essential for capturing the residual correlations among the newly generated sites that are not described by the Gaussian-mixture prior.

\emph{CNF only, without the GMM prior.}
With a sufficiently large number of ODE steps, the CNF can learn most of the required transformation and provides the dominant contribution to the model accuracy. However, the ESS decreases sharply as the number of ODE steps is reduced. In contrast, the full model maintains an ESS  approximately $86\%$ down to 15 steps and  $56\%$ at 10 steps of ode. This shows that the conditional GMM provides an effective prior for the CNF and stabilizes the model when a coarse ODE discretization is used.

The benefit of GMM becomes considerably larger as the hierarchy deepens, at $L=64$, the efficiency decreases from $0.71$ with the conditional GMM to $0.50$ without it as shown in Table \ref{tab:gmm_ode}. The difference also grows as the ODE integration budget is reduced. At $L=32$, for example, decreasing the number of steps from $20$ to $10$ retains an efficiency of $0.56$ with the GMM, whereas the efficiency without it falls to $0.09$. The conditional prior therefore enables efficient training at a substantially lower integration cost.

Conceptually, the unconditional Gaussian is farther from the target distribution of the newly introduced sites conditioned on the coarse field. Each flow must consequently perform a larger and more difficult transport, increasing the possibility of deviating from the desired conditional structure and destabilizing training. By initializing the new sites closer to their target conditional distribution, the conditional GMM reduces the transport required at each refinement level. Its advantage is therefore most pronounced for larger lattices and smaller ODE budgets, where it improves both sampling efficiency and training stability.
\begin{figure*}[t]
\centering
\makebox[\textwidth][l]{\hspace*{-0.35cm}\includegraphics[width=\textwidth]{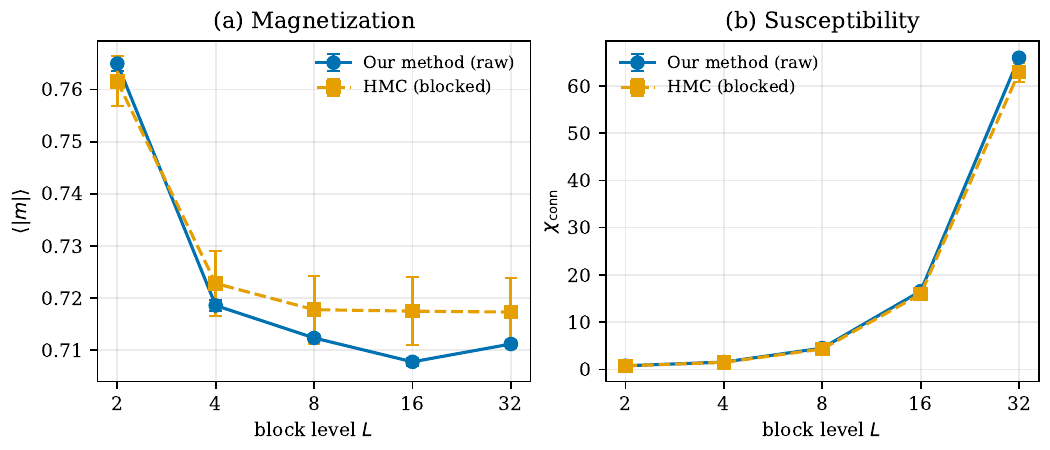}}
\caption{Comparison of the learned coarse marginals with even--even decimated HMC ensembles at $\kappa_c$ for $L_0=2$. Raw model samples and blocked HMC results are shown at $L\in{2,4,8,16,32}$ for (a)~$\langle|m|\rangle$ and (b)~$\chi_{\rm conn}$. Agreement is observed at all levels without importance reweighting.}

\label{fig:rg}
\end{figure*}

\emph{Masking:}
Replacing the two parity-restricted flows by a single flow acting on the entire fine lattice destabilizes training. The efficiency initially increases but subsequently collapses irreversibly [Fig.~\ref{fig:ablation}(d)], with a peak below $\sim0.2$. Preserving the inherited coarse sites exactly is therefore crucial for maintaining long-range correlations in the hierarchy and ensuring stable training.

\emph{Coarse lattice $L_0$.}
Varying the initial coarse resolution over $L_0\in{2,4,8}$ changes $\mathrm{ESS}/N$ only from approximately $0.81$ to $0.83$ [Fig.~\ref{fig:ablation}(b)]. Thus, within this range, the long-wavelength modes are captured comparably well and the final efficiency depends only weakly on the choice of $L_0$.

\emph{Fine ODE steps:}
The efficiency saturates at approximately $15$ integration steps per flow [Fig.~\ref{fig:ablation}(c)]. Increasing the budget beyond this value produces little additional improvement, whereas reducing it to $10$ steps lowers the efficiency and can destabilize training, as indicated by the superscript $^{*}$.

Overall, the dominant performance gain comes from combining the coarse-to-fine hierarchy with masked flow-based refinement. The precise coarse resolution is comparatively unimportant, while the conditional prior plays a more subtle but essential role in reducing the integration cost and stabilizing training.

\begin{table}[t]
\caption{Importance-sampling efficiency $\mathrm{ESS}/N$ at
$\kappa_c=0.2705$ with and without the conditional GMM prior.
The lattice-size comparison is performed using $20$ ODE steps per flow,
while the ODE-step comparison is performed at $L=32$.}
\label{tab:gmm_ode}
\begin{ruledtabular}
\begin{tabular}{lcc}
\textrm{Configuration} & \textrm{with GMM} & \textrm{without GMM} \\
\hline
\multicolumn{3}{l}{(a) Lattice size at 20 ODE steps per flow}\\
\hspace{1em}$L=32$ & 0.86 & 0.83 \\
\hspace{1em}$L=64$ & 0.71 & 0.50 \\
\hline
\multicolumn{3}{l}{(b) ODE steps per flow at $L=32$}\\
\hspace{1em}5  & 0.02 & 0.02 \\
\hspace{1em}10 & 0.56 & 0.09 \\
\hspace{1em}15 & 0.86 & 0.50 \\
\hspace{1em}20 & 0.86 & 0.82 \\
\end{tabular}
\end{ruledtabular}
\end{table}


\section{Relation to renormalization-group blocking}
\label{sec:rg}

Our construction follows a Kadanoff blocking coarse-to-fine hierarchy, but the continuous normalizing flows used at each level are not themselves RG transformations. A Wilsonian RG transformation integrates out short-distance degrees of freedom and produces an effective action with scale-dependent couplings. In contrast, our CNFs transport samples while the original fine-lattice action remains fixed. Rather than constructing the generally intractable conditional densities
$p(\phi^{(\ell)}|\phi^{(\leq \ell-1)})$, we approximate them with learned conditional models $q_\ell$, consisting of a conditional GMM prior followed by a masked CNF.

The masking structure nevertheless provides a natural RG-inspired diagnostic. Because the inherited coarse sites are preserved exactly during refinement, the restriction
\begin{equation}
  R\bigl(\phi^{(\leq \ell)}\bigr)=\phi^{(\leq \ell-1)}
\end{equation}
is deterministic and exact. The marginal distribution produced by the model at each coarse level can therefore be compared directly with an HMC ensemble obtained by applying the same restriction to fine-lattice configurations. This comparison requires neither an explicit coarse action nor a parametrization of effective couplings.

We perform this test at $\kappa_c$ using the model with $L_0=2$, whose hierarchy contains $L\in\{2,4,8,16,32\}$. Starting from an HMC ensemble at $L=32$, we repeatedly retain the even--even sublattice, $32\to16\to8\to4\to2$, matching exactly the decimation used by the model. At every level, we compare the raw, unreweighted model samples with the corresponding decimated HMC ensemble using the absolute magnetization $\langle |m|\rangle$ and the connected susceptibility
\begin{equation}
  \chi_{\rm conn}=L^2\bigl(\langle m^2\rangle-\langle |m|\rangle^2\bigr).
\end{equation}
Uncertainties are estimated by bootstrap resampling for our method and by blocked jackknife for HMC.

As shown in Fig.~\ref{fig:rg}, the model reproduces both observables across all five resolutions. The values of $\langle |m|\rangle$ agree within their combined uncertainties, while $\chi_{\rm conn}$ agrees to within a few percent. At $L=32$, for example, we obtain $\langle |m|\rangle=0.7112\pm0.0008$ compared with $0.7173\pm0.0066$ from HMC, and $\chi_{\rm conn}=66.0\pm0.3$ compared with $62.9\pm2.0$. Similar agreement is observed at every coarser level.

The small systematic offset in the raw model results---slightly lower $\langle |m|\rangle$ and slightly higher $\chi_{\rm conn}$---reflects the remaining mismatch between the learned distribution and the target, consistent with $\mathrm{ESS}/N\simeq0.81$. We deliberately use unreweighted samples here so that the agreement is not enforced by importance reweighting. The comparison therefore shows directly that the learned hierarchy produces coarse marginals consistent with genuine decimations of the fine-lattice target ensemble.

Determining the effective action associated with these coarse marginals would require a separate inverse-RG analysis. The exact blocked action is generally nonlocal and generates an increasingly large operator basis, as indicated in Eq.~\eqref{eq:rg_step}. Reconstructing its couplings and following their evolution across scales is beyond the scope of this work, whose objective is an efficient and asymptotically unbiased sampler.

\section{Training Efficiency: ESS vs.\ GPU Time}
\label{app:training_efficiency}

\begin{figure}[h]
  \centering
  \includegraphics[width=\linewidth]{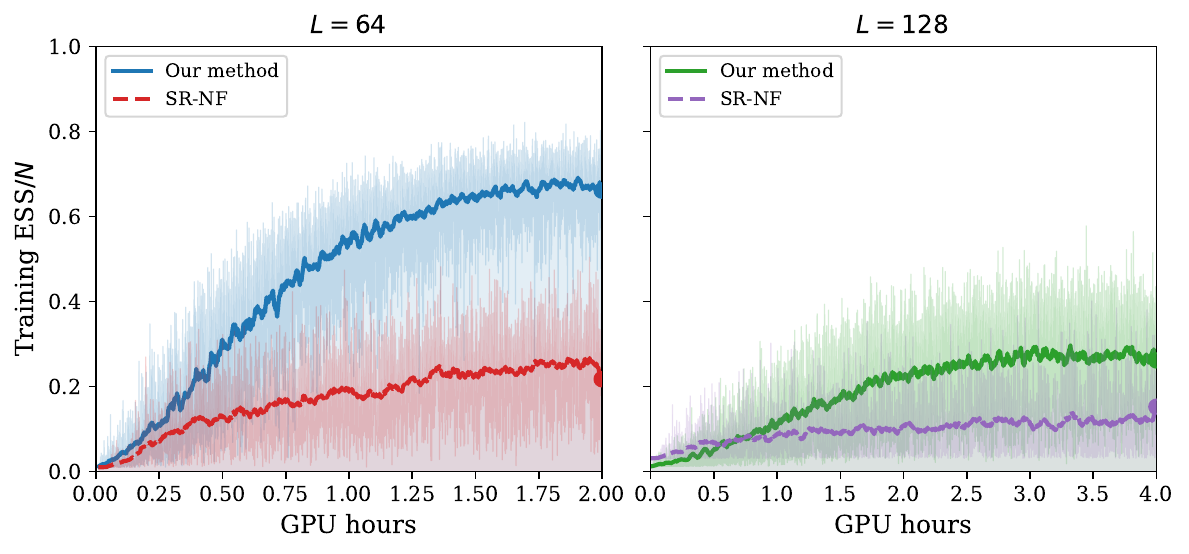}
  \caption{%
    Training ESS$/N$ (Moving Average-20) vs.\ GPU hours, clipped to 4\,hr, for
    our method (solid blue) and SR-NF (dashed orange) at $L=64$
    (left) and $L=128$ (right).
    Filled circles mark the ESS at the end of the clipped window.
  }
  \label{fig:training_gpu_hours}
\end{figure}

Figure~\ref{fig:training_gpu_hours} shows the evolution of the training
ESS$/N$ (20-epoch moving average) as a function of wall-clock GPU time,
clipped to a 4-hour budget, for our method and SR-NF at $L=64$
and $L=128$.

For $L=64$, our method completes its full training schedule ($10{,}000$ epochs)
in $\approx\!2.3$\,GPU-hours, reaching ESS$/N\approx0.67$. In contrast,
SR-NF requires $\approx\!5.3$\,GPU-hours for the same number of epochs
and achieves only ESS$/N\approx0.29$ at the 4-hour mark.

For $L=128$, the training time gap widens: our method processes each epoch roughly
$2.3\times$ faster than SR-NF.
Within 4\,GPU-hours, our method achieves ESS$/N\approx0.30$, exceeding SR-NF's value of $\approx0.13$,
with a steeper improvement trajectory.
The full run of our method ($20{,}000$ epochs, $7.8$\,GPU-hours) reaches
ESS$/N \approx 0.34$, whereas SR-NF at its full run
($20{,}000$ epochs, $18.3$\,GPU-hours) plateaus at $\approx0.15$.

\section{Computational Scaling Analysis}
\label{app:scaling}

Let $N=L^2$ denote the number of lattice sites and $B$ the batch size.
We analyze the peak training memory and explain why the baseline
methods fail to scale to large lattices.

In common implementations, 
all models are trained by
backpropagating through the full ODE integration trajectory, rather
than using the reverse integration method. As a result, the complete computation
graph over all evaluations of the ODE functions must be retained in memory
during training. For an RK4 integrator with $n_{\rm steps}$ steps,
this corresponds to storing activations from evaluations of functions $4n_{\rm steps}$.

\subsection{Dense CNF}

The Dense CNF vector field is a dense operator that is translationally
and $D_4$-equivariant acting on the entire lattice.
With the FFT implementation~\cite{gerdes2023learning}, the number of
independent parameters scales as $\mathcal{O}(N/8)$ due to orbit
reduction, while the forward cost per ODE evaluation scales as
\begin{equation}
  \mathcal{C}_{\rm Dense}^{\rm fwd}
  \sim \mathcal{O}(F_{\rm basis}\,N\log N),
\end{equation}
where
\begin{equation}
  F_{\rm basis}
  = n_{\rm sin}\times n_{\rm time}
  = 49\times 11
  = 539.
\end{equation}
The divergence is available in closed form and is therefore relatively
cheap; the dominant bottleneck is memory.
At each ODE evaluation the model materializes a basis-expansion tensor
of shape
\begin{equation}
  [B,\,N,\,F_{\rm basis}],
\end{equation}
so the peak training memory scales as
\begin{equation}
  M_{\rm Dense}
  \sim 4\,n_{\rm steps}\times B\times N\times F_{\rm basis}
  \times \mathrm{sizeof(float32)},
\end{equation}
where the leading factor of $4$ accounts for the four stages of the
RK4 integrator, each of which must retain its activations for
backpropagation.
For $L=64$ ($N=4096$, $B=64$, $n_{\rm steps}=50$), this gives
\begin{equation}
  4\times 50\times 64\times 4096\times 539\times 4\,\mathrm{B}
  \approx 113\,\mathrm{GB},
\end{equation}
which already exceeds the 80\,GB hardware limit and explains why, at $L=64$,
Dense CNF had to be trained with batch size $B=64$, whereas the hierarchical methods
use $B=100$ at the same lattice size Table~\ref{tab:arch_params}. The memory cost of exact
divergence evaluation prevents Dense CNF from reaching the same batch size.
At $L=128$ ($N=16384$), the lattice volume quadruples, driving the memory
requirement to approximately $450\,\mathrm{GB}$, far beyond available hardware.
Thus the effective scaling limitation of Dense CNF is
\begin{equation}
  M_{\rm Dense}
  \sim \mathcal{O}(F_{\rm basis}\,n_{\rm steps}\,B\,N),
\end{equation}
with a very large prefactor that makes the method memory-infeasible
beyond $L=64$ even at reduced batch sizes.
\subsection{Hutch CNF}

Hutch CNF replaces the exact divergence by a stochastic estimator. The
vector field is a 3-layer convolutional network with parameter count
independent of $L$, and the forward cost per ODE evaluation scales as
\begin{equation}
  \mathcal{C}_{\rm Hutch}^{\rm fwd}
  \sim \mathcal{O}(DHK^2N),
\end{equation}
which is formally well behaved in lattice volume.

However, the stochastic divergence estimator requires differentiating
through an additional Jacobian-based quantity at every ODE evaluation.
This introduces a substantial memory overhead beyond the forward pass.
The peak training memory therefore scales approximately as
\begin{equation}
  M_{\rm Hutch}
  \sim 4n_{\rm steps}\times B\times N\times
  \bigl(F_{\rm basis}+D\times H\bigr)\times c_{\rm div},
\end{equation}
where $c_{\rm div}>1$ represents the extra graph-storage cost of the
stochastic divergence evaluation. In practice, despite the smaller
parameter count, this makes the peak training memory comparable to
Dense CNF: $L=16$ remains feasible on a 40\,GB GPU, but already
$L=32$ requires 80\,GB-class hardware.

The second limitation is statistical rather than memory-related. The
stochastic divergence estimate has variance that grows with system
size,
\begin{equation}
  \mathrm{Var}\!\left[\widehat{\mathrm{div}}(v)\right]
  \propto N^2 \propto L^4,
\end{equation}
so the gradient signal-to-noise ratio deteriorates rapidly with $L$.
The combined effect of increased memory use and increasingly noisy
training limits Hutch CNF to small lattices.
Both SR-NF and our method are hierarchical models based on convolutional CNF components and have the same asymptotic O(L
2
) memory scaling
\subsection{Empirical observation and discussion on SR-NF vs.\ Our Method}
\label{app:emp_srfvsour}
Both SR-NF and our method are hierarchical models based on convolutional CNF components and the
same asymptotic memory scaling $\mathcal{O}(L^2)$. The difference is
therefore not the formal scaling law but the amount of work performed
per level and the resulting statistical efficiency.

At level $\ell$, our method transports only the newly introduced
sites, i.e.\ $
  \frac{3}{4}N_\ell$,
whereas SR-NF transports all
$ N_\ell$
sites. In addition, SR-NF need $70$--$100$ ODE steps per level for stable training,
while for our method $20$ ODE steps are sufficient. As a result, SR-NF is several times
more expensive per refinement level, although both methods have the
same asymptotic $\mathcal{O}(L^2)$ scaling.

The more important difference is statistical. In our method, the
masked CNF leaves the coarse sites unchanged, so the restriction
$R(\phi^{(\le \ell)}) = \phi^{(\le \ell-1)}$ is exact. Each refinement level therefore only has to learn the
conditional distribution of the newly introduced sites given the
coarse field. The associated per-level log-weight variance $ \sigma_\ell^2$
remains controlled as $L$ grows.

In SR-NF, by contrast, the full-lattice CNF updates all sites at
every level, including the coarse ones. Each level must
therefore both correct the newly introduced variables and re-correct
the coarse field. This makes the task progressively harder as the
lattice becomes larger, so that $
  \sigma_\ell^2 $
increases with $L_\ell$. Since the overall IS efficiency behaves as
\begin{equation}
  \frac{\mathrm{ESS}}{N}
  \propto
  \exp\!\left(
    -\frac{1}{2}\sum_\ell \sigma_\ell^2
  \right),
\end{equation}
the accumulated variance across levels leads to an exponential
deterioration of ESS with depth. Furthermore, unlike our method with the GMM prior, local correlation must be learned independently for each site, which can be achieved by CNF but with less statistical efficiency. This factor leads to the
observed growth of the induced autocorrelation time from $
  \tau_{\rm int}=186 \quad (L=128)$
to
$
  \tau_{\rm int}=3615 \quad (L=256).$
  
Also as in SR-NF single-step upsampler is used, we find that it is difficult to stabilise the training with a small number of ODE steps. The ablation studies in Appendix.~\ref{sec:ablation} show that the conditional GMM prior plays an important role in stabilising this training. By placing the newly introduced sites close to their conditional target distributions before the flow acts, the GMM prior enables stable and efficient sampling with a relatively small ODE integration budget.
In summary, Dense CNF fails because of the large memory footprint of
its dense exact-divergence architecture, Hutch CNF fails because of
both memory overhead and rapidly growing gradient noise, and SR-NF
fails because its unmasked hierarchical transport and lacks of systemical structure capturing local conditions across lattice sites causes strong
deterioration in ESS. Our method is the only one that avoids
all three failure modes simultaneously in the tested regime.

\section{MLMC Estimator}
\label{app:SR-NF_mlmc}
\begin{table}[h]
  \centering
  \caption{%
    Per sample cost $c_\ell$ ($\mu$s/sample at level $l$) and optimal sample count
    $N_\ell$ [Eq.~\eqref{eq:optimal_n}] for MLMC at equal
    wall-clock budget $T = N_{\rm IS}\times c_L$ with $N_{\rm IS}=10^5$.
  }
  \label{tab:mlmc_levels}
  \begin{ruledtabular}
  \begin{tabular}{ccccc}
    $L_\ell$ & \multicolumn{2}{c}{$L=32$ run} & \multicolumn{2}{c}{$L=64$ run} \\
             & $c_\ell$ ($\mu$s) & $N_\ell$ & $c_\ell$ ($\mu$s) & $N_\ell$ \\
    \hline
     2  &  13.9  & $4.76\times10^6$ &   6.5   & $6.85\times10^6$ \\
     4  &  34.4  & $2.46\times10^6$ &  17.8   & $3.51\times10^6$ \\
     8  & 135.3  & $6.32\times10^5$ &  73.2   & $9.08\times10^5$ \\
    16  & 406.6  & $1.53\times10^5$ & 238.2   & $2.23\times10^5$ \\
    32  & 6286   & $5.24\times10^4$ & 1003.6  & $5.13\times10^4$ \\
    64  & ---    & ---              & 12812   & $7.83\times10^4$ \\
  \end{tabular}
  \end{ruledtabular}
\end{table}

\begin{figure}[t]
  \centering
  \includegraphics[width=0.85\linewidth]{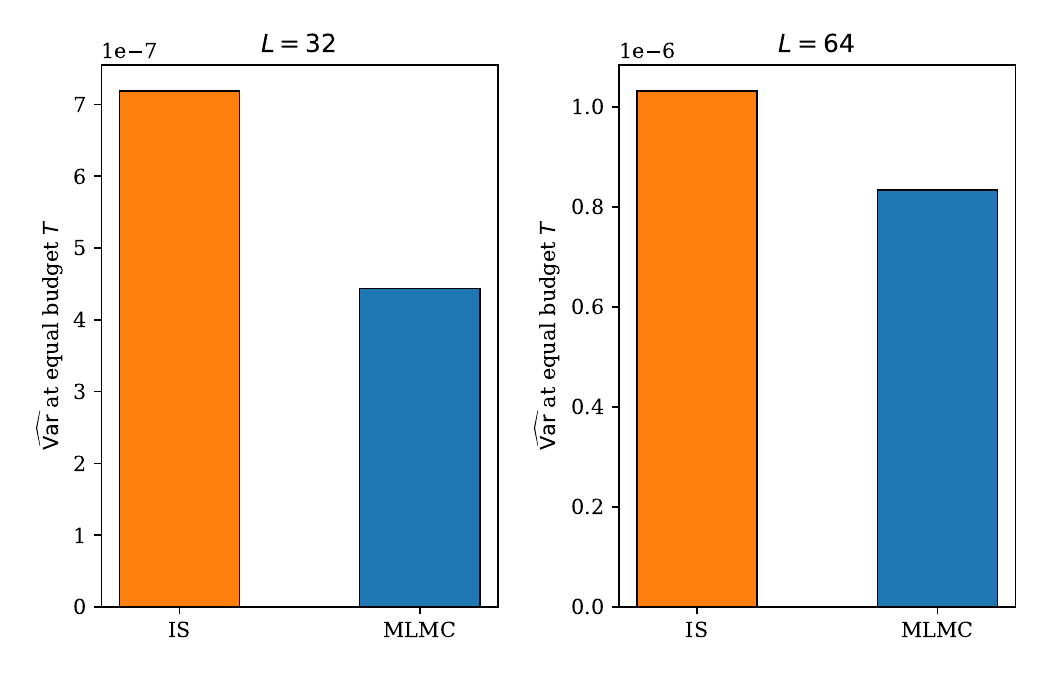}
  \caption{%
    MLMC vs plain IS variance at equal budget $T = 10^5 c_L$
    for our method at $L=32$ and $L=64$.
    MLMC (blue) achieves lower variance than IS (red) at both sizes.
    (Table~\ref{tab:mlmc}).
  }
  \label{fig:mlmc_main}
\end{figure}
\subsection{Per-Level Budget Allocation}

Table~\ref{tab:mlmc_levels} gives the per-level costs and optimal sample
counts for MLMC applied to our method at $L=32$ and $L=64$, derived from a probe run of
$10^4$ samples per level.
The coarser levels are sampled with orders-of-magnitude more configurations
than the finest level, concentrating the fine-level budget on the coarser levels which has high variance.

\subsection{Possibility to apply MLMC to SR-NF}
\label{app:SR-NF_mlmc}

MLMC requires an \emph{exact restriction map} , i.e., given a fine sample
$\phi^{(\le \ell)}$, the restriction
$R(\phi^{(\le \ell)}) = \phi^{(\le \ell-1)}$ recovers the coarse
sample actually used to generate it. For our decomposition, $R$ is 
trivial, as we preserve coarse sites at the each level.  For SR-NF this
property does not hold and the upsampling CNF acts on the full fine lattice and
displaces every site, so a naive projection of the fine output onto
the coarse sublattice yields a configuration that was never drawn from
$q^{(\le \ell-1)}$, and the level-$\ell$ MLMC correction
\begin{equation}
  \Delta^{(\ell)}_n =
  \mathcal{O}^{(\ell)}\!\bigl(\phi^{(\le \ell)}_n\bigr)
  - \mathcal{O}^{(\ell-1)}\!\bigl(\phi^{(\le \ell-1)}_n\bigr),
  \qquad
  \phi^{(\le \ell-1)}_n = R\!\left(\phi^{(\le \ell)}_n\right),
\end{equation}
is biased. The exact coarse ancestor in SR-NF can be recovered either by
integrating the level-$\ell$ CNF backwards for each fine sample or by
caching $\phi^{(\le \ell-1)}_n$ during forward generation; the former
costs one extra full-lattice ODE per level and eliminates any MLMC
advantage on its own, while the latter trades memory for compute and
is the variant we adopt below.

\begin{figure}[h]
  \centering
  \includegraphics[width=0.6\linewidth]{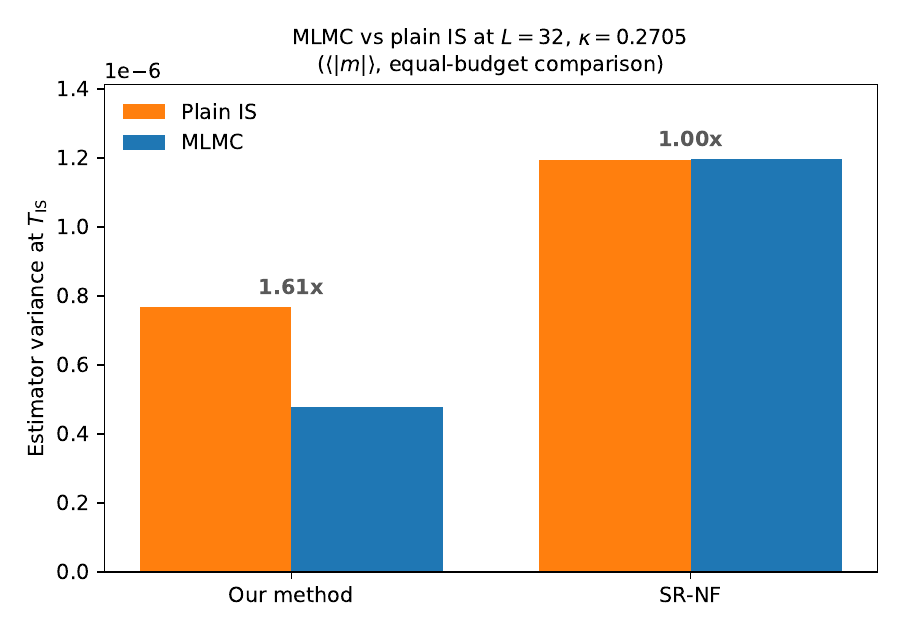}
  \caption{MLMC versus plain IS at $L=32$, $\kappa=0.2705$, for
  $\langle|m|\rangle$, at equal wall-clock budget $T_{\mathrm{IS}}$.
  Both estimators are asymptotically unbiased. SR-NF offers no computational benefit over plain IS, while
 MLMC applied to our method reduces the compute required for equivalent precision by
  roughly $38\%$.}
  \label{fig:mlmc_srnf_L32}
\end{figure}

However, unbiasedness alone is not enough. MLMC delivers a speedup
only when $\operatorname{Var}[\mathcal{O}^{(\ell)} -
\mathcal{O}^{(\ell-1)}] \ll \operatorname{Var}[\mathcal{O}^{(\ell)}]$,
which requires strong cross-level correlation.
Figure~\ref{fig:mlmc_srnf_L32} compares plain IS against MLMC for
SR-NF and for our method at $L=32$, $\kappa=0.2705$, at equal
wall-clock budget $T_{\mathrm{IS}}$: both estimators are asymptotically unbiased
within one standard deviation, but SR-NF shows no computational
benefit over plain IS, whereas our method reduces the required compute
by roughly $38\%$ for the same statistical precision. A plausible
explanation is the nature of the inter-level step where in SR-NF, the
transition from $\ell-1$ to $\ell$ injects fresh Gaussian noise on the
new sites and the observables are weakly couples between adjacent levels. Our method
avoids this by construction, as coarse sites are
preserved across levels and the inter-level coupling is strong.

\end{document}